\documentstyle[aps,pre,epsf,preprint,floats]{revtex}
\begin{document}
\def\l {{\LARGE \bf \{ }}
\def\r {{\LARGE \bf \} }}
\def\B.#1{{\bbox{#1}}}
\def\BC.#1{{\bbox{\cal{#1}}}}
\title{The Scaling Structure of the Velocity Statistics in Atmospheric
Boundary Layers}
\author {Susan Kurien$^{1,2}$,
Victor S. L'vov$^{1,3}$, Itamar Procaccia$^1$ and K.R. Sreenivasan$^2$}
\address{$^1$Department of~~Chemical Physics, The Weizmann Institute of
Science, Rehovot 76100,
Israel\\
$^2$Mason Laboratory and Department of Physics, Yale University,
New Haven, CT 06520, USA\\
$^3$Institute of Automatization and Electrometry, Russian Academy of Science,
Novosibirsk 630090, Russia}
\maketitle
\begin{abstract}
The statistical objects characterizing turbulence in real turbulent flows
differ from those of the ideal homogeneous isotropic model.
They contain contributions from various
two-dimensional and three-dimensional aspects, and from the superposition
of inhomogeneous and anisotropic contributions.
We employ the recently introduced decomposition of statistical tensor
objects into irreducible representations of the SO(3)
symmetry group (characterized by $j$ and $m$ indices,
where $j=0 \dots \infty$, $-j\le m\le j$) to disentangle some of these
contributions, separating the universal
and the asymptotic from the specific aspects of the flow. The different
$j$ contributions transform differently under rotations and so form a
complete basis in which to represent the tensor objects under study.
The experimental data are recorded with
hot-wire probes placed at various heights in the atmospheric surface layer.
Time series data from single probes and
from pairs of probes are analyzed to compute the amplitudes and
exponents of different contributions to the
second order statistical objects characterized by $j=0$, $j=1$ and
$j=2$. The analysis shows the need to make a careful distinction between
long-lived quasi two-dimensional
turbulent motions (close to the ground) and relatively
short-lived three-dimensional motions. We demonstrate
that the leading scaling exponents in the three leading sectors
($j = 0, 1, 2$) appear to be different but universal, independent of the
positions of the probe, the tensorial component considered, and the
large scale properties. The measured values of the scaling exponent are
$\zeta^{(j=0)}_2=0.68 \pm 0.01$, $\zeta^{(j=1)}_2=1.0\pm 0.15$
and $\zeta^{(j=2)}_2=1.38 \pm 0.10$. We present theoretical arguments
for the values of these exponents using the Clebsch
representation of the Euler equations;
neglecting anomalous corrections, the values obtained are
2/3, 1 and 4/3 respectively. Some enigmas and questions for the future are
sketched.
\end{abstract}
\section{Introduction}
The atmospheric boundary layer is a natural laboratory of turbulence that
is unique in that it offers very high Reynolds numbers (Re). Especially
if the measurements are made during periods when mean wind speed and direction
are roughly constant, one approaches ``controlled'' conditions that are
the goals of an experiment. Students of turbulence interested in the scaling
properties, expected to be universal in the limit Re$\to \infty$,
are thus attracted to atmospheric
measurements. On the other hand the boundary layer suffers inherently
from strong inhomogeneity (explicit dependence of
the turbulence statistics on the height) which leads to strong
anisotropies such that the vertical and horizontal directions are quite
distinguishable.
In addition, one may expect the boundary layer near the ground
to exhibit large-scale quasi two-dimensional eddies whose typical
decay times and statistics may
differ significantly from the generic three-dimensional motion.
The aim of this paper is to offer systematic methods
of analysis to resolve such difficulties, leading to a useful
extraction of the universal, three-dimensional aspects of turbulence.

Fundamentally, we propose to anchor the analysis of the statistical objects
that are important in turbulence to the irreducible representations of the
SO(3) symmetry group. Although the
turbulence that we study is non-isotropic, the Navier-Stokes
equations {\em are} invariant to all
rotations. Together with incompressibility, this invariance implies that
the hierarchy of dynamical equations satisfied by the correlation
or structure functions  are also isotropic \cite{99ALP}. This symmetry has
been used in \cite{99ALP} to show that every component of the general
solution with a given
index $j$, and hence a definite behavior under rotation, has to satisfy these
equations individually, independent of other components with different
behavior under rotation. This ``foliation'' of the hierarchical equations
motivates us to expect different scaling exponents for each component
belonging to a particular $j$ sector of the SO(3) decomposition. A preliminary
test of these ideas has been reported in \cite{98ADKLPS}. The
main result of the analysis shown below, supporting the results in
\cite{98ADKLPS}, is that in each sector of the symmetry group
scaling behavior can be found with apparently universal scaling exponents.
We demonstrate below that scale-dependent correlation functions and
structure functions can be usefully represented as a sum
of contributions with increasing index $j$ characterizing the
irreducible representations of SO(3). In such a sum the
{\em coefficients} are not universal, but the scale dependence is
characterized by universal exponents. That is, a general component
$S^{\alpha\beta}(\B.R)$ of the second rank structure function,
\begin{equation}
S^{\alpha\beta}(\B.R) \equiv \langle[u^\alpha(\B.x+\B.R)-u^\alpha(\B.x)]
[u^\beta(\B.x+\B.R)-u^\beta(\B.x)]\rangle \ ,
\end{equation}
where $\langle\dots\rangle$ stands for an ensemble average, can be
usefully presented as a sum
\begin{equation}
S^{\alpha\beta}(\B.R) =\sum_{q,j,m} a_{qjm}(R) B_{qjm}^{\alpha\beta}(\hat
\B.R) \ ,
\label{fundamental}
\end{equation}
where $\B.B_{qjm}$ are the
basis functions of the SO(3) symmetry group that depend on the direction of
$\B.R$ and $a_{qjm}$ are the coefficients that depend on the magnitude of
$\B.R$ and, in general, include any scaling behavior. In other words,
in terms of a rotation operator $O_\Lambda$ which rotates space
by an arbitrary angle $\Lambda$ one writes
\begin{equation}
O_\Lambda B^{\alpha\beta}_{qjm}(\hat\B.R)=\sum_{m'=-j}^{+j}
D_{m'm}^{(j)}(\Lambda)
B_{qjm}^{\alpha\beta}(\hat\B.R) \ . \label{rotate} \end{equation}
The $(2j+1)\times(2j+1)$ matrices $D_{m'm}^{(j)}$ are the irreducible
representations of the SO(3) symmetry group.
The index $q$ is required because, in general,
there may be more than one independent basis
function with the same indices $j$ and $m$. We note that the basis
functions $\B.B_{qjm}$ depend on the unit
vector $\hat\B.R$ only, whereas the amplitude coefficients
$a_{qjm}(R)$ depend on the {\em magnitude} of $\B.R$ only. Our main point
is that amplitudes scale in the inertial range, exhibiting universal
exponents,
\begin{equation}
a_{qjm}(R) \propto R^{\zeta_2^{(j)}} \ . \label{uniscale} \end{equation}
Analyzed by usual log-log plots, a superposition
such as (\ref{fundamental}) may well result in
continuously changing slopes, as if there is no scaling. One of our main
aims is to stress that the scaling exists, but
needs to be revealed by unfolding the various contributions.
This approach flushes out the scaling behavior even when the
Reynolds number Re is low, as in numerical simulations
\cite{99ABMP}.

Obviously, to isolate tensorial components belonging to sectors
other than the isotropic one needs to collect data from more than one probe.
In Section 2 we present the
experimental configuration and the conditions of measurement,
and discuss the nature of the
data sets. We demonstrate there that having {\em two} probes is actually
sufficient to read surprisingly rich information about anisotropic turbulence.
We have so far used two types of
geometries, one consisting of two probes at the same height above the
ground and the other with two probes vertically
separated. In both cases the inter-probe separation is orthogonal to the
mean wind. The caveat is that we must rely on Taylor's
hypothesis \cite{38Tay} to generate scale-dependent structure functions.
In anisotropic flows the validity and optimal
use of this method require discussion, and this is done in Section 3.
In that Section we also examine the issues concerning
2- and 3-dimensional aspects of the flow pattern, and
determine the outer scale $L$ at which three-dimensional scaling behavior
ceases to exist.
In Section 4 we present the main results of the analysis. We demonstrate
that the second-order structure function is best described
as a superposition of contributions belonging to different sectors of
the SO(3) symmetry group by extracting the coefficients
and exponents that appear in the superposition
(\ref{fundamental}).
The scaling exponents depend on $j$, and we will demonstrate that they are
an increasing function of $j$. For $j=0,1,2$ our data analysis leads to the
exponent values $0.68\pm 0.01$, $1.0\pm 0.15$ and $1.38\pm
0.10$ respectively. In Section 5 we present theoretical considerations
that determine these exponents neglecting intermittency
corrections, based on the Clebsch representation of the Euler equation.
Section 6 offers a summary of the principal conclusions,
and a discussion of the road ahead.
\section{The experimental setup}
\begin{figure}
\epsfxsize=14 truecm
\epsfbox{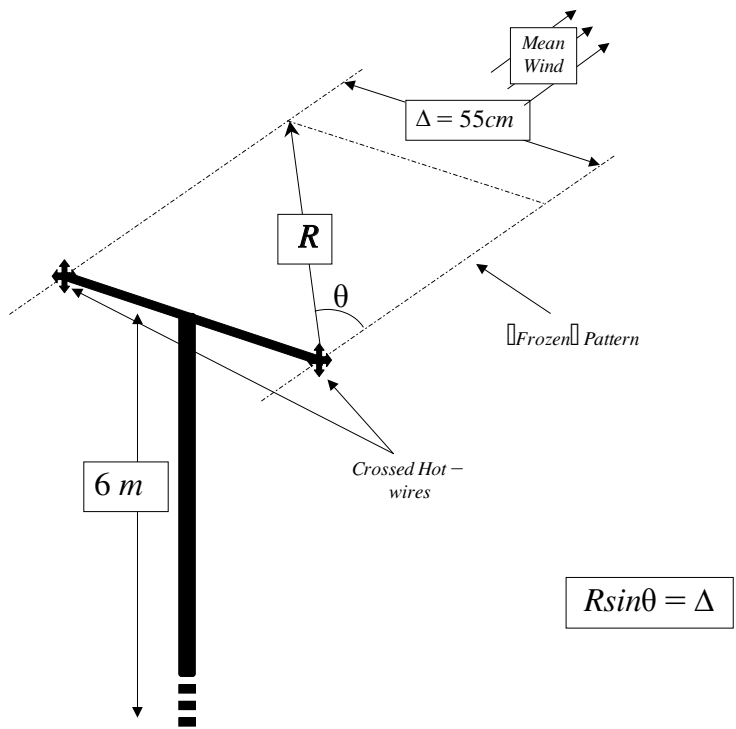}
\caption{Schematic of the experimental set-up. Shown is the
positioning of the probes with respect to the mean wind,
and an explanation of how Taylor's hypothesis
is employed.}
\label{Figexpset} \end{figure}

The results presented in this paper are based on two experimental setups,
which are denoted throughout as {\rm I} and {\rm II} respectively.
In both setups the data were acquired over the salt flats in
Utah with a long fetch. The site of measurements was chosen to provide
steady wind conditions. The surface of the desert was very smooth and even.
The measurements were made between 6~PM and 9~PM in a summer season
during which
nearly neutral stability conditions prevailed. The boundary layer was
very similar to that on a smooth flat plate.
In set {\rm I} the data were acquired
simultaneously from two single hot-wire probes at a height of
6 m above the ground, with a horizontal separation of 55 cm,
nominally orthogonal to the mean wind, see Fig.~\ref{Figexpset}.

 The Taylor
microscale Reynolds number was about 10,000. Set {\rm II} was acquired from
an array of three cross-wires, arranged {\em above} each
other at heights 11 cm, 27 cm and 54 cm respectively. The Taylor
microscale Reynolds numbers for this set were 900, 1400
and 2100 respectively. The hot-wires, about 0.7 mm in length and
6 $\mu$m in diameter, were calibrated just prior to
mounting them on the mounting posts and checked
immediately after dismounting. The hot-wires were operated on DISA 55M01
constant-temperature anemometers. The frequency
response of the hot-wires was typically good up to 20 kHz. The voltages
from the anemometers were suitably low-pass
filtered and digitized. The voltages were constantly monitored on an
oscilloscope to ensure that they did not exceed the
digitizer limits. The Kolmogorov scales were about 0.75
mm (set I) and 0.5-0.7 mm (set II). Table~I lists a
few relevant facts about the data records analyzed here. The various
symbols have the following meanings: $\overline U$ =
local mean velocity, $u^{\prime}$ = root-mean-square velocity,
$\langle \varepsilon \rangle$ = energy dissipation rate
obtained by the assumption of local isotropy and Taylor's hypothesis,
$\eta$ and $\lambda$ are the Kolmogorov scale and Taylor
microscale respectively, the microscale Reynolds number
$R_{\lambda} \equiv u^{\prime} \lambda/\nu$, and $f_s$ is the
sampling frequency.
\begin{table}
\begin{tabular} {|c|c|c|c|c|c|c|c|c|}
Height&$\overline U$ & $u^\prime$ &$10^2 \langle \varepsilon \rangle $,& $\eta$
& $\lambda$ & $R_{\lambda}$ & $f_s,$ per & \# of \\meters& ms$^{-1}$ & ms
$^{-1}$
& m $^2$ s$^{-3}$ & mm & cm & & channel, Hz & samples\\
\hline 6 & 4.1 & 1.08 & $1.1$ & 0.75 & 15 & 10,500 & 10,000 & $4 \times
10^7$\\
\hline\hline 0.11 & 2.7
& 0.47 & $6.6 $ & 0.47 & 2.8 & 900 & 5,000 & $ 8 \times 10^6$\\
0.27 &3.1 & 0.48 & 2.8& 0.6 & 4.4 & 1400& 5,000 &$8 \times 10^6$\\
0.54 &3.5 &0.5& 1.5& 0.7& 6.2&2100& 5,000& $8 \times
10^6$\\
\end{tabular}
\caption{Data sets I (first line) and II (second-fourth lines).}
\end{table}
For set I we need to test whether the separation between the two probes is
indeed
orthogonal to the mean wind. (We do not need to worry about this point in
set {\rm II},
since the probes are above each other). To do so we computed the
cross-correlation function $\langle
u_1(t+\tau)u_2(t)\rangle$. Here, $u_1$ and $u_2$ refer to velocity
fluctuations in the direction of the mean wind,
for probes 1 and 2 respectively. If the separation were precisely orthogonal
to the mean wind, this quantity should be maximum for $\tau=0$. Instead,
for set I, we
found the maximum shifted slightly
to $\tau=0.022$ s, implying that the separation was not precisely orthogonal
to the mean wind. To correct for this effect,
the data from the second probe were time-shifted by 0.022 s. This amounts
to a change in the actual value of the
orthogonal distance. We computed this effective distance to be $\Delta
\approx 54$ cm (instead of the 55 cm that was set
physically). We choose coordinates such that the mean wind direction is
along the 3-axis,
the vertical is in along the 1-axis and the third direction orthogonal to
these is the 2-axis.
We denote these directions by the three unit vectors $\hat \B.n$, $\hat \B.m$,
and $\hat \B.p$ respectively. The raw data available from set I is
$u^{(3)}(t)$ measured at the
positions of the two probes. In set II each probe reads a linear
combination of $u^{(3)}(t)$ and
$u^{(1)}(t)$ from which each component is extractable. From these data
we would like to compute the scale-dependent structure functions,
using Taylor's hypothesis to surrogate space for time.
This needs a careful discussion, which is given below.
\section{Theoretical constructs:
Taylor's Hypothesis, Inner and Outer Scales}
\subsection{Taylor's Hypothesis}
Decades of research on the statistical aspects of hydrodynamic turbulence
are based on Taylor's hypothesis [4-7], which asserts that the
fluctuating velocity field measured by a given
probe as a function of time, $\B.u(t)$, is the same as the velocity
$\B.u(R/\overline{U})$ where $\overline{U}$ is the mean
velocity and $R=-\overline{U}t$ is the distance to a position
``upstream'' where the
velocity is measured at $t=0$. The natural
limitation on Taylor's hypothesis is provided by the typical decay
time of fluctuations of scale $R$. Within the classical
scaling theory of Kolmogorov, this time scale is the turn-over
time $R/\sqrt{S(R)}$ where $S(R)\equiv S^{\alpha\alpha}(R)$. With this
estimate, Taylor's hypothesis is expected to be valid when
$\sqrt{S(R)}/\overline{U} \to 0$. Since $S(R)\to 0$ when $R\to
0$, the hypothesis becomes exact in this limit. We will use
this aspect to match the units while
reading a distance from a combination of space and time
intervals.

Reference \cite{99LPP} presents a detailed analysis of the consequences
of Taylor's hypothesis on the basis of an exactly soluble model.
It also proposes ways for
minimizing the systematic errors introduced by the use of
Taylor's hypothesis. In light of that analysis we will use an
``effective" wind, $U_{\rm eff}$, for surrogating the time data. This velocity
is a combination of the mean wind $\overline{U}$ and the
root-mean-square $u'$,
\begin{equation} U_{\rm eff}
\equiv \sqrt{\overline{U}^2+(bu')^2} \ , \label{defUeff}
\end{equation}
where $b$ is a dimensionless parameter. Evidently, when we employ the
Taylor hypothesis in log-log plots of structure functions using time
series measured in a {\em single} probe, the value of the parameter $b$
is irrelevant, because it merely changes the (arbitrary) units of length
(i.e, yields an arbitrary intercept).
When we mix real distance between two probes and
surrogated distance according to Taylor's hypothesis, the parameter
$b$ becomes a unit fixer. The numerical value of this  parameter is
found in \cite{99LPP} by the requirement that the surrogated and
directly measured structure functions
coincide in the limit $R\to 0$. When we do not have the
necessary data we will use values of $b$ suggested by the exactly
soluble model treated in \cite{99LPP}. This value of $b\approx 3$.
We have checked  that the scaling exponents change by no more than a few
percent upon changing $b$ by 30$\%$.
Further, this choice can be justified {\em a posteriori} by the
quality of the fit of to the predicted scaling functions.

When we have two probes placed at different heights, the mean velocities
and $u'$ as measured by the probes do not coincide.
In applying Taylor's hypothesis
one needs to decide the  most appropriate value of $U_{\rm eff}$.
 This question has been addressed in detail in
Ref.\cite{99LPP} with the final conclusion that the
choice depends on the velocity profile between the probe. In the case of
{\em linear}
shear the answer is
\begin{equation} U_{\rm eff}
\equiv \sqrt{{\overline{U_1}^2+\overline{U_2}\over
2}^2+b{{u_1'}^2 +{u_1'}^2\over2}} \ , \label{defUeff2}
\end{equation}
where the subscripts 1,2 refer to the two probes respectively.

\begin{figure}
\epsfxsize=16truecm
\epsfysize=7truecm
\epsfbox{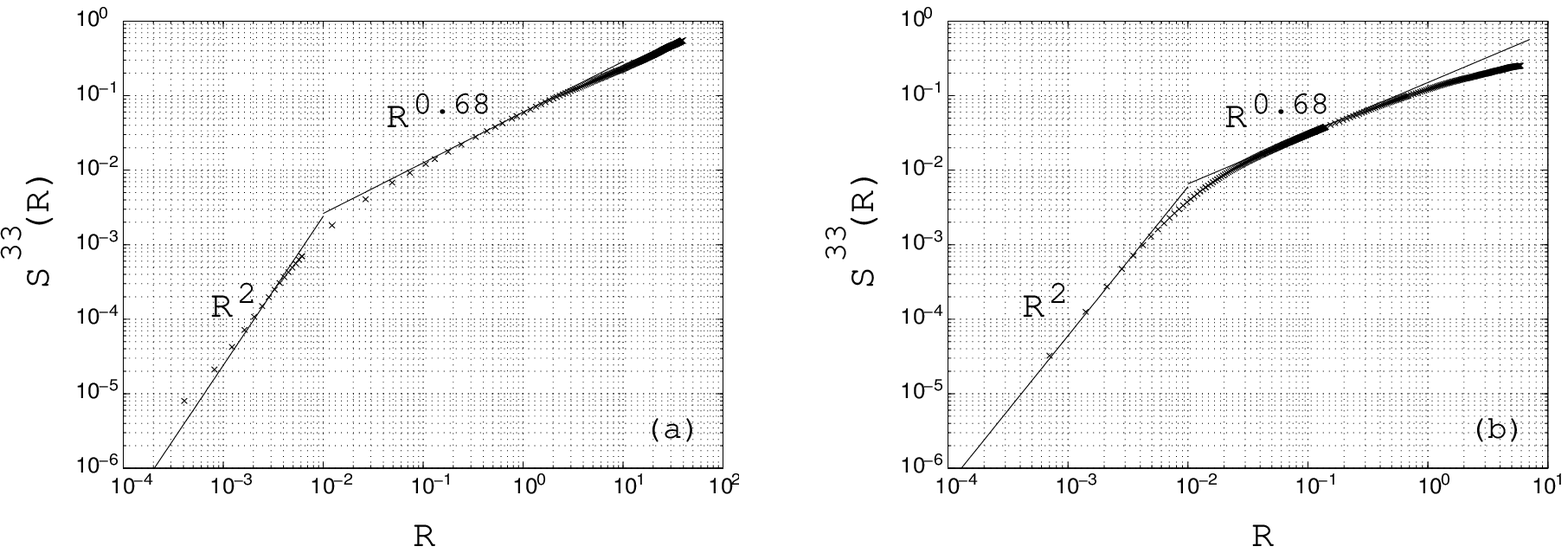}
\caption{Log-log plots of the longitudinal component of the
second order structure function. Panel~(a) is for data set I and
panel~(b) for data set II.}
\label{susanpaper.scrange_long.eps} \end{figure}

In all subsequent expressions, we will therefore denote separations by $R$,
and invariably this will mean Taylor-surrogated time differences or a
combination of real and Taylor-surrogated distances.
The effective velocity will be (\ref{defUeff}) or
(\ref{defUeff2}) depending on whether the probes are at the same height
or at different heights.
\subsection{Inner and Outer Scales in the Atmospheric Boundary Layer}
\begin{figure}
\epsfxsize=7.5truecm
\epsfysize=7truecm
\epsfbox{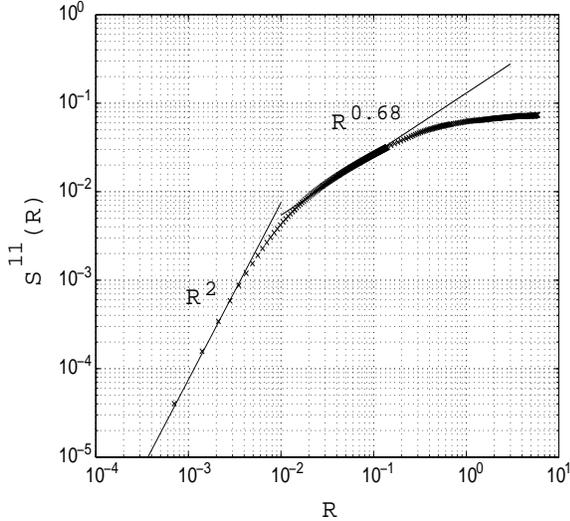}
\caption{Log-log plot of the transverse component of the
second order structure function computed from data set II.}
\label{susanpaper.scrange_trans.eps} \end{figure}

In seeking scaling behavior one needs to find the inner and outer scales.
Below the inner scale second order structure functions
have an analytic dependence on the separation, $S(R)\sim R^2$, and above
the outer scale they should
tend to a constant value. We show in Fig.~\ref{susanpaper.scrange_long.eps}
the longitudinal structure functions
\begin{equation}
S^{33}(R) = \langle(u^{(3)}(x + R) - u^{(3)}(x))^2
\rangle \end{equation}
computed from a single probe in set I and from the probe at $0.54$~m in set II.
We also consider in Fig.~\ref{susanpaper.scrange_trans.eps}
the transverse structure function
\begin{equation}
S^{11}(R) = \langle(u^{(1)}(x + R) - u^{(1)}(x))^2
\rangle \end{equation}
computed from the probe at $0.54$~m in
set II, see Fig.~\ref{susanpaper.scrange_trans.eps}.
The spatial scales are computed
using the local mean wind in both cases since we do not expect the scaling
exponent for the single-probe structure
function to be affected by the choice of advection velocity.
However, this choice does determine the value of $R$ corresponding to
a particular time scale, but we expect that any correction to
the numerical value of $R$ is small for a different
choice of advection velocity, and not crucial for the qualitative
statements that follow. In Fig.~\ref{susanpaper.scrange_long.eps}
we clearly see the $R^2$ behavior characterizing the
transition from the dissipative to the inertial range. As is well-known
\cite{Fri}, this behavior persists for
about a half-decade above the ``nominal'' Kolmogorov length scale $\eta$.
There is a region of cross-over and then the isotropic scaling
$\sim R^{0.68}$ expected for small scales in the inertial
range begins. We thus have no difficulty in identifying the inner
scale, it is simply revealed as a natural crossover length
in these data.

Next, since  we cannot expect to fit
with a single power-law for larger scales and must include scaling
contributions due to anisotropy \cite{98ADKLPS},
we need to estimate the largest
scales that should be included in our fitting procedure using the SO(3)
decomposition machinery. We expect that the contributions due to
anisotropy will account for scaling behavior up to the outer scale
of the three-dimensional flow patterns. The task now is to
identify that scale.
One approach is simply to use the
scale where the  longitudinal structure function tends to
 a constant, corresponding to
the scale across which the velocity signal has decorrelated. It becomes
immediately apparent that this is not a reasonable
estimate of the large scale. Figure~\ref{susanpaper.scrange_long.eps}
shows that the longitudinal structure function stays correlated
up to scales that are at least an order of magnitude larger
than the height at which the measurement is made.
On the other hand, the transverse structure function computed from the
probe at $0.54$~m, Fig.~\ref{susanpaper.scrange_trans.eps},
ceases to exhibit scaling behavior at a scale that is of the
order of the vertical distance of the probe from the ground.

It appears that we are observing extremely flat
 ``quasi-two dimensional''  eddies that are correlated over
very long distances in the horizontal direction but have a
comparatively small vertical velocity component. Accordingly
the vertical velocity component is dominated by {\em bona fide}
three-dimensional turbulence. Since we know that the presence
of the boundary must limit the size of the largest
three-dimensional structures, the height of the probe should be
something of an upper bound on the largest three-dimensional flow
patterns that can be detected in experiments. Thus the size of the largest
three-dimensional structures is more accurately determined by the
decorrelation length of the transverse structure function.
The theory of scaling behavior in three-dimensional turbulence can usefully be
applied to only those flow patterns that are
essentially three-dimensional. The extended flat eddies must be
described in terms of a separate theory, including notions of
two-dimensional turbulence which has very different scaling properties
\cite{67Kra}.
This is {\em not} the ambition of the present work. Rather, in the
following analyses, we choose our outer-scale $L$ in the horizontal
direction to be of the order of the decorrelation length of the
{\em transverse} structure function (where available)
or of the height of the probe. We will see below that these two are the same
to within a factor of 2; taking $L$ to be as twice the height of the probe
is consistent with all our data. We use this estimate in
our study of both transverse, longitudinal and mixed objects.
\section{Extracting the universal exponents of higher $j$
sectors}
In this section we describe a procedure for extracting the
scaling exponents that appear in the superposition (\ref{fundamental}).
Preliminary results on the scaling exponent
$\zeta_2^{(j=2)}$, obtained under the assumption of cylindrical symmetry,
were announced in \cite{98ADKLPS}. The analysis here is more
complete, and takes into account the full tensorial structure.
We show that taking into account the full broken symmetry is feasible,
and the final results are essentially the
same. Both sets of results are also in agreement with analysis of numerical
simulations \cite{99ABMP}. The results concerning $\zeta_2^{(j=1)}$ are
new.

In order to extract a particular $j$ contribution and the associated
scaling exponent, one would ideally like to possess the statistics of
the velocity at all points in a three-dimensional grid.
One could then extract the $j$ contribution of particular interest
by multiplying the full structure function by the appropriate $B_{q,j,m}$ and
integrating over a sphere of radius $R$. Orthogonality of the basis functions
ensures that only the $j$ contribution survives the integration.
One could then perform this procedure for various $R$ and extract
the scaling behavior. This method was adopted successfully
in \cite{99ABMP} using data from direct numerical simulations. The
experimental data are limited to a few points in space, so the
integration over the sphere is not possible.
We are faced with a true superposition of contributions from various $j$
sectors with no simple way of disentangling them. However we can do the
next best thing and use the postulate that the scaling exponents
form a hierarchy of increasing values for increasing $j$. This can be
interpreted to mean that anisotropic effects appear to increase
with increasing scale. Since we look for the lowest order anisotropic
contributions in our analyses, we perform a two-stage procedure to separate
the various sectors. First we look at the small scale region of the inertial
range to determine the extent of the fit with a single (isotropic) exponent.
We then seek to extend this range by including appropriate anisotropic tensor
contributions, and obtain the additional scaling exponents using
least-squares fitting procedure. The following two sections discuss
the procedure for determining the
$j=2$ and $j=1$ scaling contributions to second-order statistics.

\subsection{The $j$=2 component}
In the second order structure function defined already, {\em viz.},
\begin{equation}\label{Sab}
S^{\alpha\beta}({\B.R}) = \langle (u^\alpha({\B.x} + {\B.R}) -
u^\alpha({\B.x}))
(u^\beta({\B.x} + {\B.R}) - u^\beta({\B.x}))\rangle , \end{equation}
the $j=2$ component of the SO(3) symmetry group
corresponds to the lowest order anisotropic contribution that is
symmetric in the indices, and has even parity
in ${\B.R}$ (due to homogeneity). Although the assumption of
axisymmetry used in \cite{98ADKLPS}
seemed to be justified from the excellent qualities of fits
obtained, we attempt to fit the same data (set I) with the {\em full}
tensor form for the $j=2$ contribution. The
derivation of the full $j=2$ contribution to the symmetric, even
parity, structure function appears in Appendix A.

\begin{figure}
\epsfxsize=16.5truecm
\epsfysize=6truecm
\epsfbox{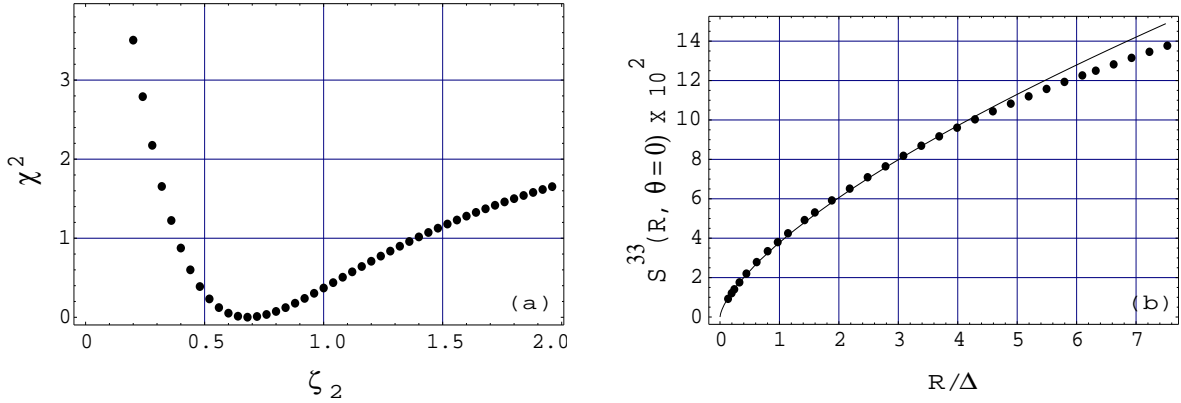}
\caption{The structure function computed from the single probe data, set I.
(a) shows the $\chi^2$ minimization by the best-fit value of the exponent
in the isotropic sector $\zeta_2\approx 0.68$ for the
single-probe structure function in the range $0.2 < R/\Delta <4.5$.
(b) shows the fit using the best value of $\zeta_2$ obtained
in (a), indicating the peel-off from isotropic  behavior at the end of the
fitted range.}\label{Fig1}
\end{figure}

To begin with, we seek the range over which the isotropic scaling
exponent holds for data set I. We measure all separation distances
in units of $\Delta = 0.54$~m which is the distance between the probes.
The lower bound to the inertial range in this set is estimated to begin
at $R/\Delta \approx 0.2$ (see discussion in Section III B).
We then find the range of scales over which the structure function
\begin{equation}\label{tta0}
S^{33}(R,\theta = 0) = \langle (u_1^{(3)}(x+R) - u_1^{(3)}(x))^2
\rangle,
\end{equation}
with the subscript $1$ denoting one of the two probes, can
be fitted with a single exponent; this then would indicates the limit
of isotropic scaling. We find that for  points in the range
$0.2<R/\Delta<4.5$ a least squares fitting procedure yields
a best-fit value $\zeta_2=0.68\pm0.01$ (Fig.~\ref{Fig1}~(a)).
Fig.~\ref{Fig1}~(b) shows the fit to the structure function computed from a
single probe in set I  with just the $j=0$
contribution. Above this range, we are unable to obtain a good fit to
the data with just the isotropic exponent
and Fig.~\ref{Fig1}~(b) shows the peel-off from isotropic behavior above
$R/\Delta \approx 4.5$. We point out that for ranges higher than this,
one can indeed able to find a ``best-fit'' exponent for the curve, but
the value of the exponent rapidly decreases and the quality of the fit is
compromised.

To find the $j=2$ anisotropic exponent we need to use data taken
from both probes. To clarify the procedure, we show in
Fig.~\ref{Figexpset} the geometry of set I. What is computed is actually
\begin{equation}
S^{33}(R,\theta)=\langle [u^{(3)}_1( U_{\rm eff} t + U_{\rm eff}t_{\tilde R})-
u^{(3)}_2( U_{\rm eff}t)]^2\rangle \end{equation}
Here $\theta=\arctan(\Delta/ U_{\rm eff}t_{\tilde R})$, $t_{\tilde R}=\tilde R/
U_{\rm eff}$,
and $R=\sqrt{\Delta^2+(\bar U_{\rm eff}t_{\tilde R})^2}$. $U_{\rm eff}$ is
defined
by Eq.~(\ref{defUeff}) with $b=3$. For simplicity we shall refer from now
on to such quantities as
\begin{equation}\label{ttadep}
S^{33}(R,\theta) = \langle (u_1^{(3)}(x+R) - u_2^{(3)}(x))^2\rangle \ .
\end{equation}

Next, we fix the scaling exponent of the isotropic sector as $0.68$ and
find the
$j=2$ anisotropic exponent that results from fitting to the full $j=2$ tensor
contribution. We fit the objects in Eqs.~(\ref{tta0}) and (\ref{ttadep})
to the sum of the $j=0$
(with scaling exponent $\zeta_2 = 0.68$) and the $j=2$ contributions (see
Appendix A)
\begin{eqnarray}
&&S^{33}(R,\theta)=S^{33}_{j=0}(R,\theta)+ S^{33}_{j=2}(R,\theta) \nonumber\\
&=&c_0\left({R\over \Delta}\right)^{\zeta_2} \Big[ 2
+\zeta_2-\zeta_2 \cos^2\theta\Big] \nonumber\\
&+&a\left({R\over \Delta}\right)^{\zeta_2^{(2)}}
\Big[ (\zeta_2^{(2)}+2)^2 -\zeta_2^{(2)}
(3\zeta_2^{(2)}+2)\cos^2\theta\nonumber\\
&+&2\zeta_2^{(2)}(\zeta_2^{(2)}-2)\cos^4\theta\Big] \nonumber\\
&+&b\left({R\over
\Delta}\right)^{\zeta_2^{(2)}} \Big[ (\zeta_2^{(2)}+2) (\zeta_2^{(2)}+3)-
\zeta_2^{(2)}(3\zeta_2^{(2)}+4)\cos^2\theta\nonumber\\
&+&(2\zeta_2^{( 2)}+1) (\zeta_2^{(2)}-2)\cos^4\theta\Big] \nonumber \\
&+&a_{9,2,1} \left({R\over \Delta}\right)^{\zeta_2^{(2)}}
\Big[-2\zeta_2^{(2)} (\zeta_2^{(2)}+2)
\sin\theta\cos\theta
+ 2\zeta_2^{(2)}(\zeta_2^{(2)}-2)\cos^3\theta\sin\theta \Big]\nonumber\\
&+&a_{9,2,2} \left({R\over
\Delta}\right)^{\zeta_2^{(2)}}\Big[-2\zeta_2^{(2)} (\zeta_2^{(2)}-2)
\cos^2\theta\sin^2\theta\Big]
\nonumber\\
&+&a_{1,2,2}\left({R\over \Delta}\right)^{\zeta_2^{(2)}}
\Big[-2\zeta_2^{(2)} (\zeta_2^{(2)}-2)\sin^2\theta\Big].
\label{fulltens}
\end{eqnarray}

\begin{figure}
\epsfxsize=7truecm
\epsfysize=5.5truecm
\epsfbox{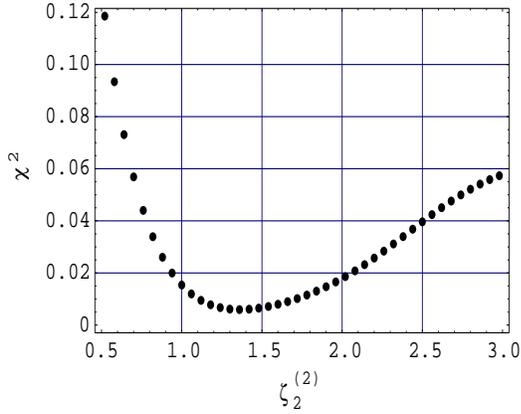}
\caption{The $\chi^2$ minimization by the best-fit value of the exponent
in the $j=2$ anisotropic sector from the fit to both the $\theta=0$ and the
$\theta$-dependent structure functions in the ranges $0.2 < R/\Delta <25$
and $1 < R/\Delta < 25$ respectively.}\label{chj2}
\end{figure}
\begin{figure}
\epsfxsize=16.5truecm
\epsfysize=6truecm
\epsfbox{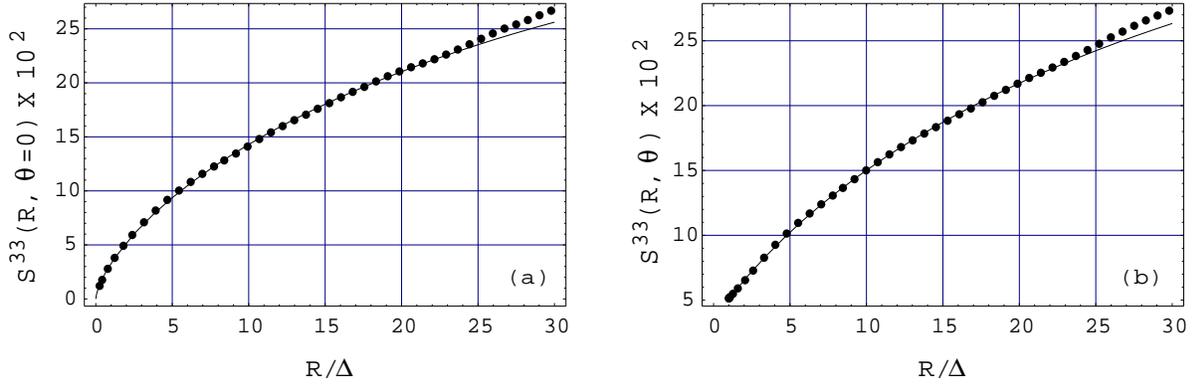}
\caption{The structure functions computed >from data set I and fit with the
$j=0$ and full $j=2$ tensor contributions using the best fit values
of exponents $\zeta_2=0.68$ and $\zeta_2^{(2)}=1.38$.
Panel (a) shows the fit to
the single-probe ($\theta = 0$) structure function in the range
$0.2 < R/\Delta < 25$ and panel (b) shows the fit to the
$\theta$-dependent structure function in the range $1 < R/\Delta < 25$.}
\label{sfj2full}
\end{figure}

We fit the experimentally generated functions to the above form
using values of $\zeta_2^{(2)}$ ranging from $0.5$ to $3$.
Each iteration of the fitting procedure involves solving for the six
unknown, non-universal coefficients. The best value of $\zeta_2^{(2)}$
is the one that minimizes the $\chi^2$ for these fits;
from Fig.~\ref{chj2} we obtain this to be $1.38\pm0.15$. The fits with
this choice of exponent are displayed in
Fig.~\ref{sfj2full}. The corresponding values of the six fitted coefficients
is given in Table~II. The range of scales that are
fitted to this expression is $0.2 < R/\Delta < 25$ for the $\theta=0$
(single-probe) structure function and $1 < R/\Delta < 25$ for the
$\theta \ne 0$ (two-probe) structure function.
We are unable to fit with Eq.~(\ref{fulltens}) to larger scales ---
that is, larger than about 12 meters ---
without losing the quality of the fit in the small scales. This is
roughly twice the height of the probe from the ground. Based on the discussion
in Sect.~II C, we should be in the regime of the largest scales where the
three-dimensional theory would hold. Therefore this limit
to the fitting range is consistent with our expectations for the
maximum scale of three-dimensional turbulence. We
conclude that the structure functions exhibits scaling behavior
over the whole scaling range, but this important fact is missed if
one does not consider a superposition of the $j=0$ and
$j=2$ contributions.
\begin{table}
\begin{tabular} {|c|c|c|c|c|c|c|c|}
$\zeta_2$ & $\zeta_2^{(2)}$ & $c_0 \times 10^3 $ & $a \times 10^3$ & $b
\times 10^3 $ & $a_{9,2,1} \times 10^3 $ & $a_{9,2,2} \times 10^3 $ &
$a_{1,2,2} \times 10^3$\\
\hline
0.68 & 1.38 $\pm 0.10$ & 7 $\pm 0.5$ & -3.2 $\pm$ 0.3 & 2.6 $\pm $0.3 &
-0.14$ \pm$ 0.02 & -5.6 $\pm $ 0.7& -4 $\pm $ 0.5
\end{tabular}
\caption{The scaling exponents and the 6 coefficients in units of (m/sec)$^2$
as determined from the nonlinear fit of Eq. 7 to data set I.}
\end{table}
We thus conclude that the estimate for the
$j=2$ scaling exponent $\zeta_2^{(2)}\approx 1.38$. This same estimate
was obtained in \cite{98ADKLPS} using only the axisymmetric terms. The
value of the coefficients $a$ and $b$ are again close
in magnitude but opposite in sign --- just as in \cite{98ADKLPS}
giving a small contribution to
$S^{33}(R,\theta=0)$. The non-axisymmetric contributions
vanish in the case of $\theta = 0$. The contribution of these terms to the
finite $\theta$ function is relatively small because the angular
dependence appears as $\sin\theta$ and $\sin^2\theta$, both of
which are small for small $\theta$ (large $R$); and hence previously
we were able to obtain a good fit to just the axisymmetric
contribution. Lastly, we note that
the total number of free parameters
in this fit is 7 (6 coefficients and 1 exponent). This
brings up the possibility of having ``over-fit'' the data. The relative
``flatness'' of the $\chi^2$ function near its minimum in
Fig.~\ref{chj2} could be indicative of the large number of free parameters
in the fit. However, the value of the exponent is perfectly in
agreement with the analysis of numerical simulations
\cite{99ABMP} in which one can properly integrate the structure function
against the basis functions, eliminating all contributions
except that of the $j=2$ sector. Furthermore, fits to the data in the
vicinity of $\zeta_2^{(2)} = 1.38$ show enough divergence from experiment
that we are satisfied about the genuineness of the $\chi^2$ result.
\subsection{Extracting the j=1 component}
The homogeneous structure function defined in Eq.~(\ref{Sab})
is known from properties of
symmetry and parity to possess no contribution from the $j=1$ sector (see
Appendix B, Section 2), the $j=2$ sector being its lowest order
anisotropic contributor. In order to isolate the scaling behavior of
the $j=1$ contribution in atmospheric shear flows we must either
explicitly construct a new tensor object which will allow for such
a contribution, or extract it from  the structure function
itself computed in the presence of {\em inhomogeneity}.
In the former case, we construct the tensor
\begin{equation}\label{Tab}
T^{\alpha\beta}({\B. R}) = \langle u^\alpha({\B. x} + {\B. R}) -
u^\alpha({\B. x}))(u^\beta({\B. x} + {\B. R}) + u^\beta({\B. x}))\rangle.
\end{equation}
This object vanishes both when $\alpha=\beta$ and when ${\B. R}$ is
in the direction of homogeneity.
>From data set II we can calculate this function for
non-homogeneous scale-separations (in the shear direction). In general,
this will exhibit mixed parity and symmetry; we cannot use the
incompressibility condition to reduce our parameter space.
Therefore, to minimize the final number of fitting parameters,
we examine only the antisymmetric contribution. We derive the tensor
contributions in the $j=1$ sector
for the antisymmetric case in Appendix B, Section 1, and use this to
fit for the unknown $j=1$ exponent. We describe the
results of this analysis below. For completeness, we have derived the
tensor contributions in the $j=1$ sector for the
symmetric case as well in Appendix B2. This can be used to find $j=1$
exponent for the  inhomogenous structure function which is symmetric but has
mixed parity. We do not present the results of that analysis here
essentially because they are consistent with those from the antisymmetric
case.
\begin{figure}
\epsfxsize=8truecm
\epsfysize=5truecm
\epsfbox{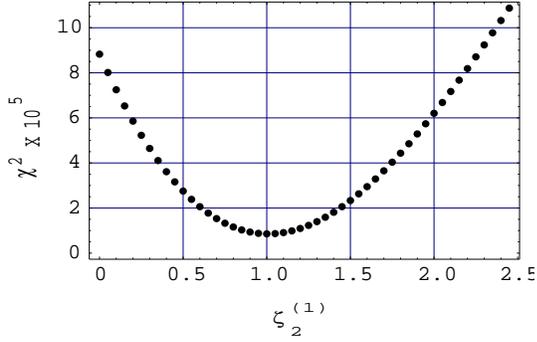}
\caption{The $\chi^2$ minimization by the best-fit value of the exponent
$\zeta_2^{(1)}$ of the $j=1$ anisotropic sector from the fit to
$\theta$-dependent ${\widetilde T}^{31}(R,\theta)$ function in
the range $1 < R/\Delta < 2.2$. }\label{chj1T} \end{figure}
\begin{figure}
\epsfxsize=8truecm
\epsfysize=6truecm
\epsfbox{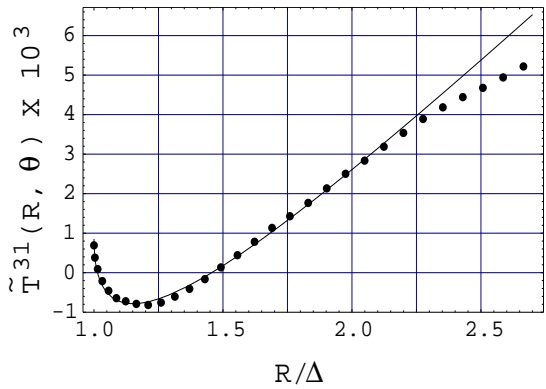}
\caption{The fitted ${\widetilde T}^{31}(R,\theta)$ function.
The dots indicate the data and the line is the fit.} \label{Tfit}
\end{figure}
Returning now to consideration of the antisymmetric part of the tensor object
defined in Eq.~(\ref{Tab}), {\em viz.},
\begin{equation}
{\widetilde T}^{\alpha\beta}({\B. R}) =
{T^{\alpha\beta}({\B. R}) - T^{\beta\alpha}({\B. R}) \over 2} =
\langle u^\alpha({\B. x})u^\beta({\B. x} + {\B. R})\rangle
- \langle u^\beta({\B. x})u^\alpha({\B. x} + {\B. R})\rangle,
\end{equation}
which will only have contributions from the antisymmetric $j=1$
basis tensors. An additional useful property of
this object is that, it does not have any contribution from the
isotropic $j=0$ sector
spanned by $\delta^{\alpha\beta}$ and $R^\alpha R^\beta$.
This allows us to isolate the $j=1$ contribution and determine its scaling
exponent $\zeta_2^{(1)}$ starting from the smallest
scales available. Using data (set II) from the probes at
$0.27$~m (probe 1) and at $0.11$~m (probe 2) we calculate
\begin{equation}
{\widetilde T}^{31}({\B. R})
= \langle u_2^{(3)}(\B.x)u_1^{(1)}(\B.x + \B.R)\rangle -
\langle u_1^{(3)}(\B.x + \B.R)u_2^{(1)}(\B.x)\rangle\ ,
\end{equation}
where again superscripts denote the velocity component and subscripts
denote the probe by which this component is measured. The goal is to fit
this experimental object to the tensor form derived in Appendix B1, Eq.~(B7),
\begin{eqnarray}\label{j1tens}
{\widetilde T}^{31}(R,\theta,\phi=0) =
- a_{3,1,0}R^{\zeta_2^{(1)}}\sin\theta
+ a_{2,1,1}R^{\zeta_2^{(1)}}
+ a_{3,1,-1}R^{\zeta_2^{(1)}}\cos\theta.
\end{eqnarray}

Figure~\ref{chj1T} gives the $\chi^2$ minimization of the fit as a function of
$\zeta_2^{(1)}$. We obtain the best value to be $1 \pm 0.15$ for the final fit.
This is shown in Fig.~\ref{Tfit}.
The fit in Fig.~\ref{Tfit}
peels off at around $R/\Delta = 2$. The values of
the coefficients corresponding to the exponent $\zeta_2^{(1)}=1$ are given
in Table~III. The maximum range of scales over which
the fit works is of the order of the height of the probes from the ground,
consistent with the considerations presented earlier.
This value of the scaling exponent of the $j=1$ sector is entirely new.
Again, we have satisfied ourselves that a different value of the exponent
yields a substantially poorer fit to the data.

In Section 5 we will present theoretical
considerations to show that the value
$\zeta_2^{(1)}=1$ is predicted by a version of classical dimensional analysis.
The present findings significantly strengthen
our proposition \cite{99ALP} that the scaling exponents in the various sectors
(at least up to $j=2$) are indeed universal.
\begin{table}
\begin{tabular}{|c|c|c|c|}
$\zeta_2^{(1)}$ & $a_{3,1,0} $ & $a_{2,1,1}$ & $a_{3,1,-1}$
\\ \hline $1 \pm 0.15 $ & $ 0.0116 \pm 0.001 $ & $ 0.0124 \pm 0.001 $ &
$-0.0062 \pm 0.001$\\
\end{tabular}
\caption{The values of the exponents and coefficients (in units of
(m/sec)$^2$)
obtained from the fit to the function ${\widetilde T}^{31}(R,\theta)$.}
\end{table}
\section{Theoretical determination of $\zeta_2^{(j)}$ for $j=1,2$}

In this section we present dimensional considerations to determine
the ``classical K41" values expected of $\zeta_2^{(1)}$ and $\zeta_2^{(2)}$.
We work at the same level as the K41 approach that
yields the value $\zeta_2^{(0)}=2/3$. This is justified
since the differences between any two values $\zeta_2^{(j)}$
and $\zeta_2^{(j')}$ for $j\ne j'$ are considerably larger
than the intermittency corrections to either of
them. We note, however, that the issue of anomalous exponents in turbulence
has now multiplied several-fold, to all the $j$ sectors, in light of
the apparent universality that has unfolded in this work.

It is easiest to produce a dimensional estimate for $\zeta_2^{(2)}$.
One simply asserts \cite{lumley} that the $j=2$ contribution
is the first one appearing in
$S^{\alpha\beta}(\B.R)$  due to the existence of a shear. Since the
shear is a second rank tensor, it can appear
linearly in the $j=2$ contribution to $S^{\alpha\beta}(\B.R)$. We thus
write for any $m$,  $-j\le m\le j$,
\begin{equation}
S_{j=2}^{\alpha\beta}(\B.R) \sim T^{\alpha\beta\gamma\delta}
{\partial \bar U^\gamma\over \partial r^\delta}
f(R,\bar \epsilon) \ , \label{Sshear}
\end{equation}
Here $T^{\alpha\beta\gamma\delta}$ is a constant
dimensionless tensor made of
$\delta^{\alpha\beta}$, $R^\alpha/R$, and {\em bilinear} contributions
made of the three unit vectors
$\hat \B.p,~\hat\B.m,~\hat \B.n$, as exemplified in appendix A.
The way Eq.(\ref{Sshear}) is represented means that the dimensional
function $f(R,\bar
\epsilon)$ stands for the response of the second order structure function to a
small external shear. Ad such it is an inherent property of {\em isotropic}
turbulence. Within
the standard Kolomogorov-41 dimensional reasoning this function in the
inertial interval can be
made only of the mean energy flux per unit
time and mass, $\bar \epsilon$ and $R$ itself. The only
combination of $\bar \epsilon$ and $R$ that yields the right dimensions
of the function $f$ is $\bar \epsilon^{1/3} R^{4/3}$. Therefore
\begin{equation}
S_{j=2}^{\alpha\beta}(\B.R) \sim T^{\alpha\beta\gamma\delta}
{\partial \bar
U^\gamma\over
\partial r^\delta}
\bar \epsilon^{1/3} R^{4/3} \ . \label{yofi1}
\end{equation}
We thus find a ``classical K41"
 value of $\zeta_2^{(2)}=4/3$. Thus this
simple argument seems to rationalize nicely the experimentally found value
$\zeta_2^{(2)}=1.38\pm 0.1$.

To understand the value of $\zeta_2^{(1)}$ we cannot
proceed in the same way.
We need a contribution that is linear (rather than bilinear) in the unit
vectors $\hat \B.p,~\hat\B.m,~\hat \B.n$. We cannot
construct a contribution that is linear in the shear
and yet does not vanish due to the
incompressibility constraint. Thus there is a fundamental difference
between the $j=2$ contribution and the $j=1$ term. While the former can be
understood as an inhomogeneous term linear in the forced shear, the $j=1$ term,
being more subtle, may be  connected to a
solution of some homogeneous equation well within
the inertial interval. In fact, all the known inertial-interval
spectra in turbulent systems are related to the existence of a a flux of some
conserved  quantity which has a representation as an integral of some
density in $\B.k$-space. For example the kinetic energy may be written as
$\int d\B.k |\B.u (\B.k,t)|^2$. A well-known other integral of motion in
hydrodynamics with such a presentations
is the helicity
\begin{equation}\label{hel}
H=\int  d\B.r \,\big(\B.u \cdot \B.\nabla\times \B.u \big)\ .
\end{equation}
Thus  the helicity may be considered as a natural candidate which is
responsible for a new solution in the inertial interval that may
rationalize the $j=1$ finding. We show that this is not the case in the
following way.

The dimensionality  of $H$ (denoted as [$H$])  differs  from the
dimensionality  of
the energy $E$ by one length: $[H]=[E/R]$. Correspondingly, the
dimensionality  of the
helicity flux, $\bar h$ may be written as
\begin{equation}\label{hel-fl}
[\bar h]=[\bar \epsilon /R]\ .
\end{equation}
It means that in turbulence with energy and helicity fluxes one has at one's
disposal a dimensionless factor in the form $\bar h\, R/\bar \epsilon$. It
means
that the second order structure function $\B.S(R,\bar \epsilon ,
   \bar h   )$ cannot be found just by dimensional reasoning even within
the K41 approach. Nevertheless, assuming that   at  small  helicity fluxes
({\em i.e.} when $\bar h\, R/\bar \epsilon\ll 1$) the function
$\B.S(R,\bar \epsilon ,\bar h   )$ may be expanded in powers of   $\bar h$
 we can justify the first order correction due to helicity, $\delta_h \B.S$, as
\begin{equation}\label{hel-cor}
\delta_h  \B.S(R,\bar \epsilon ,
   \bar h   )\sim (\bar\epsilon)^{-1/3}\bar h R^{5/3} \ .
\end{equation}
The value of the inferred scaling exponent, i.e.  5/3,
is much larger than the value unity found experimentally. We thus need
to find another invariant that may rationalize the findings.

The only invariance in addition to the conservation of helicity
that we are aware of in the inviscid limit is the Kelvin circulation
theorem, which however does not furnish a local integral in $\B.k$-space
in the Eulerian representation. The only
way that is apparent to us to expose this invariance in a useful way is the
Clebsch representation,
in which one writes the Euler equation in terms of one complex field
$a(\B.r,t)$, see for example\cite{91Lvo}.   In
$\B.k$-representation the Fourier component of the velocity field
$\B.u(\B.k,t)$ is determined from
a bilinear combination of the complex field:
\begin{eqnarray}
\B.u(\B.k,t)&=&\frac{1}{8\pi^3}\int d^3k_1 d^3k_2
\B.\Psi(\B.k_1,\B.k_2)a^*(\B.k_1,t)
a(\B.k_2,t) \ , \\
\B.\Psi(\B.k_1,\B.k_2)&=&\frac{1}{2}\left(\B.k_1+\B.k_2-(\B.k_1-\B.k_2)
\frac {k_1^2-k_2^2}{|\B.k_1-\B.k_2|^2}\right) \ . \end{eqnarray}
It is well-known \cite{91Lvo}
that this representation exposes a local conserved
integral of motion which is
\begin{equation}
\B.\Pi = \frac{1}{8\pi^3}\int d^3k \B.k a^*(\B.k,t)a(\B.k,t)
\ . \end{equation}
Note that this conserved quantity is a vector, and it cannot have
a finite mean in an isotropic system.
Consider now correction  $\delta_ \B.\pi S(R,\bar\epsilon,\bar\B.\pi)$
to the second order  structure function
due to a flux $\bar\B.\pi$ of the integral of motion $\bar\B.\pi$.
The dimensionality of $\bar\B.\pi$, $[\bar\B.\pi]
=[\bar\epsilon^{2/3}/R^{1/3}]$ and therefore now the
dimesionless factor is
$\bar \B.\pi\,R^{1/3} /\bar\epsilon^{2/3}$. Assuming again expandability
of $\delta_ \B.\pi S$ at small values of the flux $\bar\B.\pi$ one finds
that
\begin{equation} S_{j=1}^{\alpha\beta}(\B.R) \sim
T^{\alpha\beta\gamma}\,\bar\pi^\gamma\, R \ , \label{yofi2}
\end{equation}
where $T^{\alpha\beta\gamma}$ is a constant dimensionless tensor {\em
linear} in
the unit vectors $\hat \B.p,~\hat\B.m,~\hat
\B.n$. We thus find the ``classical K41" value $\zeta_2^{(1)}=1$, which
should be compared with the experimental finding $\zeta_2^{(1)}=1\pm 0.15$.

We stress that Eqs. (\ref{yofi2}) and (\ref{yofi1}) are the analogs of the
standard
isotropic dimensional estimate
\begin{equation} S_{j=0}^{\alpha\alpha}(R) \sim
\left( \bar\epsilon R\right )^{2/3} \ . \label{yofi3}
\end{equation}
We thus conclude that dimensional analysis predicts that values 2/3, 1 and
4/3 for
$\zeta_2^{(j)}$ with $j=0$, 1 and 2 respectively. This appears to be
in satisfactory agreement with the experimentally extracted values of these
exponents. We should state however that we do not know at present how
to continue this line of argument for $j>2$.

\section{Summary, conclusions, and the road ahead}
In summary, we considered the second order tensor functions of velocity
in the atmospheric boundary layers. The following conclusions appear
important:
\begin{enumerate}
\item The atmospheric boundary layer exhibits three-dimensional statistical
turbulence
intermingled with activities whose statistics are quite different. The
latter are
eddies with quasi two-dimensional nature, correlated over extremely large
distances compared to the height of the measurement, having
little to do with the three-dimensional fluctuations discussed above.

\item We found that the ``outer scale of turbulence" as measured by the
three-dimensional
statistics is of the order of twice the height of the probe.

\item The inner scale is the the usual dissipative crossover, which is
clearly seen
as the scale connecting two different slopes in log-log plots.

\item Between the inner and the outer scales, Eq.~(\ref{fundamental})
appears to offer an excellent representation of the structure function.
Using contributions with
$j=0,1,2$ we could fit the whole range very accurately.

\item The scaling exponents $\zeta_2^{(j)}$ are measured as $0.68\pm 0.01,~
1\pm 0.15, ~1.38\pm0.10$ respectively.

\item Classical K41 dimensional considerations yield the numbers 2/3, 1
and 4/3 respectively. To get $\zeta_2^{(2)}=4/3$ all that we need is to
assume a contribution linear in the shear. For getting $\zeta_2^{(1)}=1$
we need to identify a non-obvious conserved quantity which allows
a new solution in the depth of the inertial interval. This is the first
time that Clebsch variables allowed the understanding of a fundamentally
new universal scaling exponent.
\end{enumerate}

If the trends seen here continue for higher $j$ values, we can rationalize the
apparent tendency towards isotropy with decreasing scales. If indeed every
anisotropic
contribution introduced by the large scale forcing (or boundary conditions)
decays
as $(R/L)^{\zeta_2^{(j)}}$ with increasing $\zeta_2^{(j)}$ as a function of
$j$,
then obviously when $R/L\to 0$ only the isotropic contribution survives.
This is a pleasing
notion that justifies the modeling of turbulence as isotropic at the small
scales.

We need to raise a word of caution here. We really have no idea
about the values of the exponents for $j\ge 3$.
Moreover, we are not even sure that they
are well defined. To understand the difficulty one needs to examine the
hierarchical equations for the correlation functions.
These equations contain integrals used to eliminate the pressure
contributions. The integrals
were proven to converge (in the IR and the UV limits) when the the exponents
$\zeta_2$ lies within the ``window of convergence" which is $(0,4/3)$
{\bf (reference?)}. We see
that with $j=3$ we may reach beyond this window of convergence (this being
questionable even for our experimental finding of $j=2$!), and we are not
guaranteed to have the kind of local theory that is thought to be a
prerequisite to scaling behavior.

Another enigma is related to the apparent success of the considerations of
Section
5 to rationalize the numerical values of the exponents found in the experiment.
There is, however, no well-defined  procedure of continuing the estimates for
$\zeta_2^{(j)}$ for $j\ge 3$. Whether this is related to the locality issue
is not understood at present.

In conclusion, it appears that we have here an exciting possibility of
generalizing the scaling structure of the statistical turbulence to
many sectors
of the symmetry group, gaining much better understanding of the structure
of a theory. There exist, however, large patches of {\em terra incognita} on
our map, patches that we hope to penetrate in future work.

\acknowledgements
At Weizmann, the work was supported by the Basic Research Fund administered
by the Israeli Academy of Sciences, the German-Israeli Foundation,
The European Commission under the TMR program, and the Naftali and Anna
Backenroth-Bronicki Fund for Research in Chaos and
Complexity. At Yale, it was supported by the National Science Foundation
grant DMR-95-29609, and the Yale-Weizmann Exchange
Program. Special thanks are due to Mr.~Brindesh Dhruva and Mr.~Christopher
White for their help in
acquiring the data.

\appendix
\section{Full form for the $j=2$ contribution for the homogeneous case}
Each index $j$ in the SO(3) decomposition of an $n$-rank tensor
labels a $2j+1$ dimensional SO(3) representation.
Each dimension is labeled by $m=-j,-j+1,\dots j$. The $j=0$ sector is the
isotropic contribution while higher order $j$'s should
describe any anisotropy. The $j=0$ terms are well-known
\begin{equation} S_{j=0}^{\alpha\beta}({\bf
R})=c_0R^{\zeta_2} \left[(2+\zeta_2) \delta^ {\alpha\beta}-\zeta_2{R^\alpha
R^\beta\over R^2}\right]\ , \label{Siso}
\end{equation} where $\zeta_2\approx0.68$ is the known universal scaling
exponent for the isotropic contribution and $c_0$ is
an unknown coefficient that depends on the boundary conditions of the flow.
For the $j=2$ sector which is the lowest
contribution to anisotropy to the homogeneous structure function, the $m=0$
(axisymmetric) terms were derived from constraints
of symmetry, even parity (because of homogeneity) and incompressibility on
the second order structure function \cite{98ADKLPS}
\begin{eqnarray}\label{m0}
&S&^{\alpha\beta}_{j=2,m=0} ({\B.R}) =
aR^{\zeta_2^{(2)}}\Big[(\zeta_2^{(2)}
-2)\delta^{\alpha\beta} -
\zeta_2^{(2)}(\zeta_2^{(2)}+6)\nonumber\\ &\times&\delta^{\alpha\beta}
{(\B.n\cdot \B.R)^2 \over R^2}+2\zeta_2^{(2)}(\zeta_2^{(2)}-2){R^\alpha
R^\beta(\B.n\cdot \B.R)^2 \over R^4}\nonumber\\
&+&([\zeta_2^{(2)}]^2+3\zeta_2^{(2)}+6)n^\alpha n^\beta \nonumber\\
&-&{\zeta_2^{(2)}(\zeta_2^{(2)}-2)\over R^2}(R^\alpha n^\beta + R^\beta
n^\alpha)(\B.n\cdot \B.R)\Big]\label{finalform}\\ &+&
bR^{\zeta_2^{(2)}}\Big [-(\zeta_2^{(2)} +3)(\zeta_2^{(2)}+2)\delta^
{\alpha\beta}(\B.n\cdot \B.R)^2 +
{R^\alpha R^\beta \over R^2} \nonumber\\ &+& (\zeta_2^{(2)} +3)
(\zeta_2^{(2)}+2)n^\alpha n^\beta + (2\zeta_2^ {(2)}+1)(\zeta_2^{(2)}-2)
\nonumber\\
&\times&{{R^\alpha}{R^\beta}{(\B.n\cdot \B.R)^2} \over R^4}-
([\zeta_2^{(2)}]^2 - 4)(R^\alpha n^\beta + R^\beta n^\alpha)(\B.n\cdot
\B.R)\Big] \ .
\nonumber
\end{eqnarray}
where $\zeta_2^{(2)}$ is the universal scaling exponent for the $j=2$
anisotropic sector and $a$ and $b$ are independent unknown coefficients to
be determined by the boundary conditions. We would
now like to derive the remaining $m=\pm1$, and $m=\pm2$ components
\begin{equation}
S_{j=2,m}^{\alpha\beta}=
	\sum_q {a_{q,2,m}R^{\zeta_2^{(2)}}B_{q,2,m}^{\alpha\beta} (\bf
{\hat R} )}
\end{equation}
where $\zeta_2^{(j=2)}$ is the scaling exponent of the $j=2$ SO(3)
representation
of the $n=2$ rank correlation function. The $B_{q,j,m}^{\alpha\beta}
(\B. {\hat R})$ are the basis functions in the SO(3) representation of the
structure function, The $q$ label denotes the
different possible ways of arriving at the the same $j$ and runs over all
such terms with the same parity and symmetry (a
consequence of homogeneity and hence the constraint of incompressibility)
\cite{99ALP}. In our case, even parity and symmetric
in the two indices. In all that follows, we work closely with the procedure
outlined in \cite{99ALP}. Following the convention
in \cite{99ALP} the $q$'s to sum over are $q=\{1,7,9,5\}$. The
incompressibility condition $\partial_\alpha u^\alpha = 0$
coupled with homogeneity can be used to give relations between the
$a_{q,j,m}$ for a given $(j,m)$. That is, for $j=2$,
$m=-2\dots 2$ \begin{eqnarray} (\zeta_2^{(2)} - 2)a_{1,2,m} +
2(\zeta_2^{(2)} - 2)a_{7,2,m} + (\zeta_2^{(2)} + 2)a_{9,2,m} &=&
0 \\ \nonumber a_{1,2,m} + (\zeta_2^{(2)} + 3)a_{7,2,m} +
\zeta_2^{(2)}a_{5,2,m} &=& 0 \end{eqnarray}

We solve equations (3) in order to obtain $a_{5,2,m}$ and $a_{7,2,m}$ in
terms of linear combinations of $a_{1,2,m}$ and $a_{9,2,m}$.
\begin{eqnarray}
a_{5,2,m} &=& {a_{1,2,m}([\zeta_2^{(2)}]^2 - \zeta_2^{(2)} - 2) + a_{9,2,m}
([\zeta_2^{(2)}]^2 + 5 \zeta_2^{(2)} + 6) \over
2\zeta_2^{(2)}(\zeta_2^{(2)} - 2)} \\ \nonumber a_{7,2,m} &=&
{a_{1,2,m}(2-\zeta_2^{(2)}) - a_{9,2,m}(2+\zeta_2^{(2)}) \over
2(\zeta_2^{(2)} - 2)}
\end{eqnarray}

Using the above constraints on the coefficients, we are now left
with a linear combination of just two linearly independent tensor forms
{\em for each m}
\begin{eqnarray}\label{genl-s2m}
S^{\alpha\beta}_{j=2,m} &=&
a_{9,2,m}R^{\zeta_2^{(2)}}[-\zeta_2^{(2)}(2+\zeta_2^{(2)})
B_{7,2,m}^{\alpha\beta}({\B. {\hat R}}) + 2\zeta_2^{(2)}(\zeta_2^{(2)} - 2)
B_{9,2,m}^{\alpha\beta}({\B. {\hat R}}) \nonumber \\
&+&([\zeta_2^{(2)}]^2+5\zeta_2^{(2)}+6)B_{5,2,m}^{\alpha\beta}({\B. {\hat
R}})] \nonumber\\
&+&a_{1,2,m}R^{\zeta_2^{(2)}}[2\zeta_2^{(2)}(\zeta_2^{(2)} - 2)
B_{1,2,m}^{\alpha\beta}({\B. {\hat R}}) -
\zeta_2^{(2)}(\zeta_2^{(2)}-2)B_{7,2,m}^{\alpha\beta}({\B. {\hat R}})
\nonumber\\
&+&([\zeta_2^{(2)}]^2-\zeta_2^{(2)}-2)B_{5,2,m}^{\alpha\beta}({\B.{\hat
R}})] \end{eqnarray}

The task remains to find the explicit form of the basis tensor functions
$B_{q,2,m}^{\alpha\beta}({\B. {\hat R}})$, $q\in\{1,7,9,5\}$,
$m\in\{\pm1,\pm2\}$
\\
$\bullet$\,$ B_{1,2,m}^{\alpha\beta}({\B. {\hat R}}) \equiv
R^{-2}\delta^{\alpha\beta}R^j Y_{2m}({\B. {\hat R}})$ \\
$\bullet$\,$ B_{7,2,m}^{\alpha\beta}({\B. {\hat R}}) \equiv R^{-2}[R^\alpha
\partial^\beta +
R^\beta \partial^\alpha]R^2 Y_{2m}({\B. {\hat R}})$ \\
$\bullet$\,$ B_{9,2,m}^{\alpha\beta}({\B. {\hat R}}) \equiv R^{-4}R^\alpha
R^\beta R^2 Y_{2m}({\B. {\hat R}})$ \\
$\bullet$\,$ B_{5,2,m}^{\alpha\beta}({\B. {\hat R}}) \equiv \partial^\alpha
\partial^\beta R^2 Y_{jm}({\B. {\hat R}})$

We obtain the $m=\{\pm1,\pm2\}$ basis functions in the following derivation.
We first note that it is more convenient to form a real basis from the $
R^2 Y_{2m}({\B. {\hat R}})$ since we ultimately wish
to fit to real quantities and extract real best-fit parameters.
We therefore form the $R^2 {\widetilde Y}_{2k}({\B. {\hat R}})$ ($k=
-1,0,1$) as follows:
\begin{eqnarray}
R^2 {\widetilde Y}_{2\; 0}(\hat {\B. R})&=&R^2 Y_{2\;0}(\hat {\B. R})
=R^2 \cos^2 \theta = R{_3}^2\nonumber \\ R^2 {\widetilde Y}_{2\;-1}(\hat
{\B. R}) &=&R^2 {Y_{2\;-1}(\hat {\B. R}) -
Y_{2\;+1}(\hat {\B. R}) \over 2}\nonumber\\ &=&R^2 {(\cos\phi -
i\sin\phi)\cos\theta\sin\theta + (\cos\phi +
i\sin\phi)\cos\theta\sin\theta \over 2} \nonumber\\ &=&R^2
\cos\theta\sin\theta\cos\phi = R_3 R_1 \nonumber\\ R^2 {\widetilde
Y}_{2\;+1}(\hat {\B. R}) &=&R^2 {Y_{2\; -1}(\hat {\B. R}) + Y_{2\; +1}(\hat
{\B. R}) \over -2i} \nonumber\\ &=&R^2 {(\cos\phi -
>i\sin\phi)\cos\theta\sin\theta - (\cos\phi + i\sin\phi)\cos\theta\sin\theta
\over -2i}\nonumber\\ &=&R^2
\cos\theta\sin\theta\sin\phi = R_3 R_2\nonumber\\ R^2 {\widetilde
Y}_{2\;-2}(\hat {\B. R}) &=& R^2 {Y_{2\; 2}(\hat {\B. R}) -
Y_{2\; -2}(\hat {\B. R}) \over 2i} \nonumber\\ &=& R^2 {(\cos2\phi +
i\sin2\phi)\sin^2\theta - (\cos2\phi -
i\sin2\phi)\sin^2\theta \over 2i} \nonumber \\ &=& R^2
\sin2\phi\sin^2\theta = 2R_1 R_2 \nonumber \\ R^2 {\widetilde
Y}_{2\;+2}(\hat {\B. R}) &=& R^2 {Y_{2\; 2}(\hat {\B. R}) + Y_{2\; -2}(\hat
{\B. R}) \over 2} \nonumber\\ &=& R^2 {(\cos2\phi +
i\sin2\phi)\sin^2\theta + (\cos2\phi - i\sin2\phi)\sin^2\theta \over 2}
\nonumber \\ &=& R^2 \cos2\phi\sin^2\theta = R_1^2 -
R_2^2 \end{eqnarray} This new basis of $R^2 {\tilde Y}_{2k}{(\B. R)}$ is
equivalent to using the $R^2 Y_{jm}{(\B. R)}$
themselves as they form a complete, orthogonal (in the new k's) set. We
omit the normalization constants for the spherical
harmonics for notational convenience. The subscripts on $R$ denote its
components along the 1 ($m$), 2 ($p$) and 3 ($n$)
directions. ${\B. m}$ denotes the shear direction, ${\B. p}$ the horizontal
direction parallel to the boundary and orthogonal
to the mean wind direction and ${\B. n}$ the direction of the mean wind.
This notation makes it simple to take the derivatives
when we form the different basis tensors and the only thing to remember is that
\begin{eqnarray}
\partial^\alpha R_1 = \partial^\alpha (\B. {R \cdot m}) = m^\alpha
\nonumber\\ \partial^\alpha R_2 = \partial^\alpha (\B. {R \cdot p})=
p^\alpha \nonumber\\ \partial^\alpha R_3 = \partial^\alpha
(\B. {R \cdot n}) = n^\alpha \end{eqnarray}

We use the above identities to proceed to derive the basis tensor functions
\begin{eqnarray}
B_{1,2, -1}^{\alpha\beta}({\B. {\hat R}}) &=& R^{-2}\delta^{\alpha\beta}
({\B. R \cdot n})({\B. R \cdot m}) \nonumber\\ B_{7,2, -1}^{\alpha\beta}
({\B. {\hat R}}) &=& R^{-2}[(R^\alpha m^\beta + R^\beta
m^\alpha)({\B. R \cdot n}) + (R^\alpha n^\beta + R^\beta n^\alpha)({\B. R
\cdot m})] \nonumber\\ B_{9,2, -1}^{\alpha\beta}({\B.
{\hat R}}) &=&R^{-2} R^\alpha R^\beta ({\B. R \cdot n})({\B. R \cdot
m})\nonumber\\ B_{5,2, -1}^{\alpha\beta}({\B. {\hat R}})
&=&n^\alpha m^\beta + n^\beta m^\alpha \nonumber\\ B_{1,2,
1}^{\alpha\beta}({\B. {\hat R}}) &=& R^{-2}\delta^{\alpha\beta}({\B.
R \cdot n})({\B. R \cdot p})\nonumber\\ B_{7,2, 1}^{\alpha\beta}({\B. {\hat
R}}) &=& R^{-2}[(R^\alpha p^\beta + R^\beta
p^\alpha)({\B. R \cdot n}) + (R^\alpha n^\beta + R^\beta n^\alpha)({\B. R
\cdot p})] \nonumber\\ B_{9,2, 1}^{\alpha\beta}({\B.
{\hat R}}) &=&R^{-2} R^\alpha R^\beta ({\B. R \cdot n})({\B. R \cdot
p})\nonumber\\ B_{5,2,1}^{\alpha\beta}({\B. {\hat R}}) &=&
n^\alpha p^\beta + n^\beta p^\alpha \nonumber\\ B_{1,2,
-2}^{\alpha\beta}({\B. {\hat R}}) &=&2 R^{-2}\delta^{\alpha\beta}({\B.
R \cdot m})({\B. R \cdot p})\nonumber\\ B_{7,2, -2}^{\alpha\beta}({\B.
{\hat R}}) &=&2 R^{-2}[(R^\alpha p^\beta + R^\beta
p^\alpha)({\B. R \cdot m}) + (R^\alpha m^\beta + R^\beta m^\alpha)({\B. R
\cdot p})]\nonumber \\ B_{9,2, -2}^{\alpha\beta}({\B.
{\hat R}}) &=& 2 R^{-2} R^\alpha R^\beta ({\B. R \cdot m})({\B. R \cdot
p})\nonumber\\ B_{5,2,-2}^{\alpha\beta}({\B. {\hat R}})
&=& 2 (m^\alpha p^\beta + m^\beta p^\alpha) \nonumber\\ B_{1,2,
2}^{\alpha\beta}({\B. {\hat R}}) &=&
R^{-2}\delta^{\alpha\beta}[({\B. R \cdot m})^2 - ({\B. R \cdot p})^2]
\nonumber\\ B_{7,2, 2}^{\alpha\beta}({\B. {\hat R}}) &=&2
R^{-2}[(R^\alpha m^\beta + R^\beta m^\alpha)({\B. R \cdot m}) - (R^\alpha
p^\beta + R^\beta p^\alpha)({\B. R \cdot
p})]\nonumber \\ B_{9,2, 2}^{\alpha\beta}({\B. {\hat R}}) &=& R^{-2}
R^\alpha R^\beta [({\B. R \cdot m})^2 - ({\B. R \cdot
p})^2] \nonumber\\ B_{5,2,2}^{\alpha\beta}({\B. {\hat R}}) &=& 2 (m^\alpha
m^\beta - p^\alpha p^\beta) \end{eqnarray}

Note that for each dimension k the tensor is bilinear in some combination
of two basis vectors from the set ${\B. m}$, ${\B. p}$ and ${\B. n}$.{bf.
can we say something here about why this is so... using shear ${\partial
u^\alpha over \partial r^\beta}$ etc...??} Substituting these tensors forms
into Eq. \ref{genl-s2m} we obtain the full tensor
forms for the $j=2$ non-axisymmetric terms, with two independent
coefficients for each k.

\begin{eqnarray}\label{ktens}
S^{\alpha\beta}_{j=2,k=-1}({\B. R})
&=&a_{9,2,-1}R^{\zeta_2^{(2)}}{\Big [}-\zeta_2^{(2)}(2+\zeta_2^{(2)})
R^{-2}[(R^\alpha m^\beta + R^\beta m^\alpha)({\B. R \cdot n})\nonumber\\
&+& (R^\alpha n^\beta + R^\beta n^\alpha)({\B. R \cdot m})] +
2\zeta_2^{(2)}(\zeta_2^{(2)} - 2)
R^{-4} R^\alpha R^\beta ({\B. R \cdot n})({\B. R \cdot m})\nonumber\\
&+&([\zeta_2^{(2)}]^2+5\zeta_2^{(2)}+6)(n^\alpha m^\beta + n^\beta
m^\alpha) {\Big ]}\nonumber\\&+&
a_{1,2,-1}R^{\zeta_2^{(2)}}{\Big [}2\zeta_2^{(2)}(\zeta_2^{(2)} - 2)
R^{-2}\delta^{\alpha\beta}({\B. R \cdot n})({\B. R \cdot m}) \nonumber\\
&-& \zeta_2^{(2)}(\zeta_2^{(2)}-2) R^{-2}[(R^\alpha m^\beta + R^\beta
m^\alpha)({\B. R \cdot n})
+ (R^\alpha n^\beta + R^\beta n^\alpha)({\B. R \cdot m})] \nonumber\\
&+&([\zeta_2^{(2)}]^2-\zeta_2^{(2)}-2)(n^\alpha m^\beta + n^\beta m^\alpha)
{\Big ]} \nonumber\\
S^{\alpha\beta}_{j=2,k=1}({\B. R})
&=&a_{9,2,1}R^{\zeta_2^{(2)}}{\Big [}-\zeta_2^{(2)}(2+\zeta_2^{(2)})
R^{-2}[(R^\alpha p^\beta + R^\beta p^\alpha)({\B. R \cdot n})\nonumber\\
&+& (R^\alpha n^\beta + R^\beta n^\alpha)({\B. R \cdot p})] +
2\zeta_2^{(2)}(\zeta_2^{(2)} - 2)
R^{-4} R^\alpha R^\beta ({\B. R \cdot n})({\B. R \cdot p})\nonumber\\
&+&([\zeta_2^{(2)}]^2+5\zeta_2^{(2)}+6)(n^\alpha p^\beta + n^\beta
p^\alpha) {\Big ]}\nonumber\\
&+&a_{1,2,1}R^{\zeta_2^{(2)}}{\Big [}2\zeta_2^{(2)}(\zeta_2^{(2)} - 2)
R^{-2}\delta^{\alpha\beta}({\B. R \cdot n})({\B. R \cdot p}) \nonumber\\
&-& \zeta_2^{(2)}(\zeta_2^{(2)}-2) R^{-2}[(R^\alpha p^\beta + R^\beta
p^\alpha) ({\B. R \cdot n}) + (R^\alpha n^\beta + R^\beta n^\alpha)({\B. R
\cdot p})] \nonumber\\
&+&([\zeta_2^{(2)}]^2-\zeta_2^{(2)}-2)(n^\alpha p^\beta + n^\beta p^\alpha)
{\Big ]}\nonumber\\
S^{\alpha\beta}_{j=2,k=-2}({\B. R})
&=&a_{9,2,-2}R^{\zeta_2^{(2)}}{\Big [}-2\zeta_2^{(2)}(2+\zeta_2^{(2)})
R^{-2}[(R^\alpha p^\beta + R^\beta p^\alpha)({\B. R \cdot m})\nonumber\\
&+& (R^\alpha m^\beta + R^\beta m^\alpha)({\B. R \cdot p})] +
2\zeta_2^{(2)}(\zeta_2^{(2)} - 2)
R^{-4} R^\alpha R^\beta ({\B. R \cdot p})({\B. R \cdot m}) \nonumber\\ &+&
([\zeta_2^{(2)}]^2+5\zeta_2^{(2)}+6)(m^\alpha p^\beta + m^\beta p^\alpha)
{\Big ]}\nonumber\\&+&
a_{1,2,-2}R^{\zeta_2^{(2)}}{\Big [}2\zeta_2^{(2)}(\zeta_2^{(2)} - 2)
R^{-2}\delta^{\alpha\beta}({\B. R \cdot m})({\B. R \cdot p}) \nonumber\\
&-&2\zeta_2^{(2)}(\zeta_2^{(2)}-2) R^{-2}[(R^\alpha p^\beta + R^\beta
p^\alpha) ({\B. R \cdot m}) + (R^\alpha m^\beta + R^\beta m^\alpha)({\B. R
\cdot p})] \nonumber\\
&+&2([\zeta_2^{(2)}]^2-\zeta_2^{(2)}-2)(m^\alpha p^\beta + m^\beta
p^\alpha) {\Big ]}\nonumber\\
S^{\alpha\beta}_{j=2,k=2}({\B. R})
&=&a_{9,2,2}R^{\zeta_2^{(2)}}{\Big [}-2\zeta_2^{(2)}(2+\zeta_2^{(2)})
R^{-2}[(R^\alpha m^\beta + R^\beta m^\alpha)({\B. R \cdot m})\nonumber\\
&-& (R^\alpha p^\beta + R^\beta p^\alpha)({\B. R \cdot p})] +
2\zeta_2^{(2)}(\zeta_2^{(2)} - 2)
R^{-4} R^\alpha R^\beta [({\B. R \cdot m})^2-({\B. R \cdot
p})^2]\nonumber\\ &+&2([\zeta_2^{(2)}]^2+5\zeta_2^{(2)}+6)(m^\alpha m^\beta
- p^\beta p^\alpha) {\Big ]}\nonumber\\&+&
a_{1,2,2}R^{\zeta_2^{(2)}}{\Big [}2\zeta_2^{(2)}(\zeta_2^{(2)} - 2)
R^{-2}\delta^{\alpha\beta}[({\B. R \cdot m})^2-({\B. R \cdot
p})^2]\nonumber\\ &-&2\zeta_2^{(2)}(\zeta_2^{(2)}-2) R^{-2}[(R^\alpha
m^\beta + R^\beta m^\alpha) ({\B. R \cdot m}) - (R^\alpha p^\beta + R^\beta
p^\alpha)({\B. R \cdot p})] \nonumber\\
&+&2([\zeta_2^{(2)}]^2-\zeta_2^{(2)}-2)(m^\alpha m^\beta - p^\beta
p^\alpha) {\Big ]}
\end{eqnarray}

Now we want to use this form to fit for the scaling exponent
$\zeta_2^{(2)}$ in the structure function
$S^{33}(\B. R)$ from data set I where $\alpha=\beta=3$ and the azimuthal
angle of $\B. R$ in the geometry is $\phi = \pi/2$.

\begin{eqnarray}
S^{33}_{j=2,k=-1}(R,\theta,\phi=\pi/2)&=&0\nonumber\\
S^{33}_{j=2,k=1}(R,\theta,\phi=\pi/2)
&=&a_{9,2,1}R^{\zeta_2^{(2)}}[-2\zeta_2^{(2)}(\zeta_2^{(2)}+2)
\sin\theta\cos\theta \nonumber \\
&+& 2\zeta_2^{(2)}(\zeta_2^{(2)}-2)\cos^3\theta\sin\theta]\nonumber \\
S^{33}_{j=2,k=-2}(R,\theta,\phi=\pi/2)&=&0\nonumber \\
S^{33}_{j=2,k=2}(R,\theta,\phi=\pi/2)
&=&a_{9,2,2}R^{\zeta_2^{(2)}}[-2\zeta_2^{(2)}(\zeta_2^{(2)}-2)
\cos^2\theta\sin^2\theta]
\nonumber\\
&+&a_{1,2,2}R^{\zeta_2^{(2)}}[-2\zeta_2^{(2)}(\zeta_2^{(2)}-2)\sin^2\theta]
\end{eqnarray}
\begin{table}
\begin{tabular}{|c|c|c|c|c|c|c|c|c|}
&\multicolumn{2}{c|}{$\phi=\pi/2,\alpha=\beta=3$} &
\multicolumn{2}{c|}{$\phi = 0,\alpha=\beta=3$} & \multicolumn{2}{c|}{$\phi
= 0,\alpha=\beta=1$}& \multicolumn{2}{c|}{$\phi = 0,\alpha=3,\beta=1$}\\
\cline {2-9} $k$&$\theta \ne 0$ &$\theta = 0$ & $\theta \ne 0$ &$ \theta =
0$ & $\theta \ne 0$ & $\theta = 0$ & $\theta \ne 0 $&$\theta = 0$ \\ \hline
0	& 2 & 2 & 2 & 2 & 2 & 2 & 2 & 0 \\ \hline
-1 & 0 & 0 & 1 & 0 & 1 & 0 & 2 & 2 \\ \hline 1	& 1 & 0 & 0 & 0 & 0 & 0 & 0
& 0 \\ \hline
-2) & 0 & 0 & 0 & 0 & 0 & 0 & 0 & 0 \\ \hline 2	& 2 & 0 & 2 & 0 & 2 & 2 & 2
& 0 \\ \hline \hline
Total	& 5 & 2 & 5 & 2 & 5 & 4 & 6 & 2 \\
\end{tabular}
\caption{The number of free coefficients in the $j=2$ sector for homogeneous
turbulence and for different geometries}
\end{table}
We see that choosing a particular geometry eliminates certain tensor
contributions.
In the case of set I we are left with 3 independent coefficients for $m\ne0$,
the 2 coefficients from the $m=0$ contribution
(Eq. \ref{m0}), and the single coefficient from the isotropic sector
\ref{Siso}, giving a total of 6 fit parameters.
The general forms in \ref{ktens} can be used along with the $k=0$
(axisymmetric)
contribution \ref{Siso} to fit to any second order tensor object.
For convenience, Table~IV
shows the number of independent coefficients that a few different experimental
geometries we have will allow in the $j=2$ sector.
It must be kept
in mind that these forms are to be used {\em only} when there is known to
be homogeneity. If
there is inhomogeneity, then we cannot apply the incompressibility
condition to provide
constraints in the various parity and symmetry sectors and we must in general
mix different parity objects, using only the
geometry of the experiment itself to eliminate any terms.
\section{The j=1 component in the inhomogeneous case}
\subsection{Antisymmetric contribution}
	We consider the tensor
\begin{equation}
T^{\alpha\beta}({\B. R}) = <u^\alpha({\B. x} + {\B. R}) - u^\alpha({\B.
x}))
(u^\beta({\B. x} + {\B. R}) + u^\beta({\B. x}))> \end{equation}.
This object is trivially zero for $\alpha=\beta$. In our experimental setup,
we measure at points separated in the shear direction and therefore have
inhomogeneity which makes the object of mixed parity
and symmetry.  We cannot apply the incompressibility condition in same
parity/symmetry sectors as before to provide
constraints. We must in general use all 7 irreducible tensor forms.
This would mean fitting $for 7 x 3 = 21$ independent
coefficients plus 1 exponent $\zeta_2^{(1)}$ in the anisotropic sector,
together with 2 coefficients in the isotropic sector. In
order to pare down the number of parameter we are fitting for, we look
at the antisymmetric part of $T^{\alpha\beta}({\B. R})$
\begin{equation}
{\widetilde T}^{\alpha\beta}({\B. R}) = {T^{\alpha\beta}({\B. R}) -
T^{\beta\alpha}({\B. R}) \over 2}
= \langle u^\alpha({\B. x})u^\beta({\B. x} + {\B. R})\rangle - \langle
u^\beta({\B. x})u^\alpha({\B. x} + {\B. R})\rangle \end{equation}
which will only have contributions from the antisymmetric $j=1$ basis
tensors. These are \\
$\bullet$ Antisymmetric, odd parity
\begin{equation}
B_{3,1,m}^{\alpha\beta}
= R^{-1}[R^\alpha\partial^\beta- R^\beta\partial^\alpha] RY_{1,m}(\B. {\hat R})
\end{equation}
$\bullet$ Antisymmetric, even parity
\begin{eqnarray}
B_{4,1,m}^{\alpha\beta} &=&
R^{-2}\epsilon^{\alpha\beta\mu}R_\mu R Y_{1,m}(\B. {\hat R}) \nonumber\\
B_{2,1,m}^{\alpha\beta} &=&
R^{-2}\epsilon^{\alpha\beta\mu}\partial_\mu R Y_{1,m}(\B. {\hat R})
\end{eqnarray}

As with the $j=2$ case we form a real basis $R {\tilde Y}_{1,k}(\B. {\hat R})$
from the (in general) complex $R Y_{1,m}(\B. {\hat R})$ in order to obtain
real coefficients in our fits.
\begin{eqnarray}
R {\tilde Y}_{1,k=0}(\B. {\hat R}) &=& R Y_{1,0}(\B. {\hat R})= R\cos\theta
= R_3 \nonumber\\
R {\tilde Y}_{1,k=1}(\B. {\hat R}) &=& R {Y_{1,1}(\B. {\hat R}) +Y_{1,1}
(\B. {\hat R}) \over 2i } \nonumber \\ &=&R\sin\theta\sin\phi = R_2
\nonumber\\ R {\tilde Y}_{1,k=-1}(\B. {\hat R}) &=& R
{Y_{1,-1}(\B. {\hat R}) - Y_{1,1}(\B. {\hat R}) \over 2 } \nonumber \\
&=&R\sin\theta\cos\phi = R_1 \nonumber\\
\end{eqnarray}
And the final forms are
\begin{eqnarray}
B_{3,1,0}^{\alpha\beta}(\B. {\hat R})
&=& R^{-1}[R^\alpha n^\beta- R^\beta n^\alpha] \nonumber\\
B_{4,1,0}^{\alpha\beta}(\B. {\hat R})
&=&R^{-2}\epsilon^{\alpha\beta\mu}R_\mu ({\B. R.n}) \nonumber\\
B_{2,1,0}^{\alpha\beta}(\B. {\hat R})
&=& R^{-2}\epsilon^{\alpha\beta\mu}n_\mu \nonumber\\
B_{3,1,1}^{\alpha\beta}(\B. {\hat R})
&=& R^{-1}[R^\alpha p^\beta- R^\beta p^\alpha] \nonumber\\
B_{4,1,1}^{\alpha\beta}(\B. {\hat R})
&=&R^{-2}\epsilon^{\alpha\beta\mu}R_\mu ({\B. R.p}) \nonumber\\
B_{2,1,1}^{\alpha\beta}(\B. {\hat R})
&=& R^{-2}\epsilon^{\alpha\beta\mu}p_\mu \nonumber\\
B_{3,1,-1}^{\alpha\beta}(\B. {\hat R})
&=& R^{-1}[R^\alpha m^\beta- R^\beta m^\alpha] \nonumber\\
B_{4,1,-1}^{\alpha\beta}(\B. {\hat R})
&=&R^{-2}\epsilon^{\alpha\beta\mu}R_\mu ({\B. R.m}) \nonumber\\
B_{2,1,-1}^{\alpha\beta}(\B. {\hat R})
&=& R^{-2}\epsilon^{\alpha\beta\mu}m_\mu \end{eqnarray}

Note: For a given k the representations is symmetric about a particular axis
in our chosen coordinate system (1=m (shear), 2=p (horizontal), 3=n
(mean-wind))

We now have 9 independent terms and we cannot apply incompressibility
in order to reduce the number of independent coefficients in our fitting
procedure. We use the geometrical constraints of our
experiment to do this.

$\bullet$ $\phi = 0$ (vertical separation), $\alpha = 3, \beta = 3$
\begin{eqnarray}
B_{3,1,0}^{31}(R,\theta,\phi=0) = -\sin\theta \nonumber
\\ B_{2,1,1}^{31}(R,\theta,\phi=0) = 1 \nonumber\\
B_{3,1,-1}^{31}(R,\theta,\phi=0) = \cos\theta \end{eqnarray}

There are no contributions from the reflection-symmetric terms
in the $j=0$ isotropic sector since these are symmetric in the indices. The
helicity term in $j=0$ also doesn't contribute
because of the geometry. So, to lowest order
\begin{eqnarray}
{\widetilde T}^{\alpha\beta}({\B. R})
&=& {\widetilde T}_{j=1}^{\alpha\beta}({\B. R}) \nonumber
\\ &=& a_{3,1,0}(R)(-\sin\theta) + a_{2,1,1}(R) + a_{3,1,-1}(R)\cos\theta
\end{eqnarray}

We have 3 unknown independent coefficients and 1 unknown exponent to fit
for in our data.

\subsection{Symmetric contribution}

	We consider the structure function
\begin{equation}
S^{\alpha\beta}({\B. R}) = <(u^\alpha({\B. x} + {\B. R}) - u^\alpha({\B.
x}))
(u^\beta({\B. x} + {\B. R}) - u^\beta({\B. x}))> \end{equation}
in the case where we have homogeneous flow. This object is symmetric in
the indices by construction, and it is easily seen that homogeneity implies
even parity in R \begin{eqnarray}
	S^{\alpha\beta}({\B. R}) &=& S^{\beta\alpha}({\B. R})\nonumber \\
S^{\alpha\beta}({\B. -R}) &=& S^{\alpha\beta}({\B. R}) \end{eqnarray}
We reason that this object
cannot exhibit a $j=1$ contribution from the $SO(3)$ representation in
the following manner. Homogeneity allows us to use the incompressibility
condition
\begin{eqnarray}
	\partial_\alpha S^{\alpha\beta} &=& 0 \nonumber\\
	\partial_\beta S^{\alpha\beta} &=& 0
\end{eqnarray}
separately on the basis tensors of a given parity and symmetry in
order to give relationships between their coefficients. For the even
parity, symmetric case we have for general $j \geq 2$
just two basis tensors and they must occur in some linear combination
with incompressibility providing a constraint between the two coefficients.
However, for $j=1$ we only have one such tensor in
the even parity, symmetric group. Therefore, by incompressibility, its
coefficient must vanish. Consequently, we cannot have a
$j=1$ contribution for the even parity (homogeneous), symmetric structure
function. 	Now, we consider the case as available in
experiment when ${\B. R}$ has some component in the inhomogeneous
direction. Now, it is no longer true that
$S^{\alpha\beta}({\B. R})$ is of even parity and moreover it is also not
possible to use incompressibility as above to exclude
the existence of a $j=1$ contribution. We must look at all $j=1$ basis
tensors that are symmetric, but not confined to even
parity. These are \\
$\bullet$ Odd parity, symmetric
\begin{eqnarray}
B_{1,1,k}^{\alpha\beta}({\B. {\hat R}})
&\equiv& R^{-1}\delta^{\alpha\beta}R {\tilde Y}_{1k}({\B. {\hat
R}})\nonumber\\ B_{7,1,k}^{\alpha\beta}({\B. {\hat R}})
&\equiv& R^{-1}[R^\alpha \partial^\beta + R^\beta \partial^\alpha] R{\tilde
Y}_{1k}(\B. {\hat R}) \nonumber \\ B_{9,1,k}^{\alpha\beta}({\B. {\hat R}})
&\equiv& R^{-3}R^\alpha R^\beta R {\tilde Y}_{1k}({\B. {\hat R}})\nonumber
\\ B_{5,1,k}^{\alpha\beta}({\B. {\hat R}})
&\equiv& R \partial^\alpha \partial^\beta R{\tilde Y}_{1k}({\B. {\hat R}})
\equiv 0
\end{eqnarray}
$\bullet$ Even parity, symmetric
\begin{eqnarray}
B_{8,1,k}^{\alpha\beta}({\B. {\hat R}})
&\equiv& R^-2[R^\alpha \epsilon^{\beta\mu\nu} R_\mu \partial_\nu + R^\beta
\epsilon^{\alpha\mu\nu} R_\mu \partial_\nu] R{\tilde Y}_{1k}(\B. {\hat R})
\nonumber\\ B_{6,1,k}^{\alpha\beta}({\B. {\hat R}})
&\equiv& [\epsilon^{\beta\mu\nu} R_\mu \partial_\nu \partial_\alpha +
\epsilon^{\beta\mu\nu} R_\mu \partial_\nu \partial_\beta] R{\tilde
Y}_{1k}(\B. {\hat R})
\equiv 0
\end{eqnarray}
We use the real basis of $R^{-1}{\tilde Y}_{1k}({\B. {\hat R}})$ which are
formed from the $R^{-1}Y_{1m}({\B. {\hat R}})$. Both
$B_{5,1,k}^{\alpha\beta}({\B. {\hat R}})$ and $B_{6,1,k}^{\alpha\beta}({\B.
{\hat R}})$ vanish because of the taking of the double derivative of an
object of single power in $R$. We thus have 4 different contributions to
symmetric $j=1$ and each of these is of 3 dimensions $(k= -1,0,1)$ giving
in general 12 terms in all. \begin{eqnarray}
B_{1,1,0}^{\alpha\beta}({\B. {\hat R}})
&=& R^{-1}\delta^{\alpha\beta}({\B. R \cdot n})\nonumber\\
B_{7,1,0}^{\alpha\beta}({\B. {\hat R}})
&=& R^{-1}[R^\alpha n^\beta + R^\beta n^\alpha] \nonumber\\
B_{9,1,0}^{\alpha\beta}({\B. {\hat R}})
&=&R^{-3}R^\alpha R^\beta ({\B. R \cdot n})\nonumber\\
B_{8,1,0}^{\alpha\beta}({\B. {\hat R}})
&\equiv& R^{-2}[(R^\alpha m^\beta + R^\beta m^\alpha)({\B. R \cdot p})
-(R^\alpha p^\beta + R^\beta p^\alpha)({\B. R \cdot m})]\nonumber \\
B_{1,1,1}^{\alpha\beta}({\B. {\hat R}})
&=& R^{-1}\delta^{\alpha\beta}({\B. R \cdot p})\nonumber\\
B_{7,1,1}^{\alpha\beta}({\B. {\hat R}})
&=& R^{-1}[R^\alpha p^\beta + R^\beta p^\alpha] \nonumber\\
B_{9,1,1}^{\alpha\beta}({\B. {\hat R}})
&=&R^{-3}R^\alpha R^\beta ({\B. R \cdot p}) \nonumber\\
B_{8,1,1}^{\alpha\beta}({\B. {\hat R}})
&\equiv& R^{-2}[(R^\alpha m^\beta + R^\beta m^\alpha)({\B. R \cdot n})
-(R^\alpha n^\beta + R^\beta n^\alpha)({\B. R \cdot m})] \nonumber \\
B_{1,1,-1}^{\alpha\beta}({\B. {\hat R}}) &=&
R^{-1}\delta^{\alpha\beta}({\B. R \cdot m})\nonumber\\
B_{7,1,-1}^{\alpha\beta}({\B. {\hat R}}) &=& R^{-1}[R^\alpha m^\beta +
R^\beta m^\alpha] \nonumber\\ B_{9,1,-1}^{\alpha\beta}({\B. {\hat R}})
&=&R^{-3}R^\alpha R^\beta ({\B. R \cdot m}) \nonumber\\
B_{8,1,-1}^{\alpha\beta}({\B. {\hat R}}) &\equiv& R^{-2}[(R^\alpha p^\beta
+ R^\beta p^\alpha)({\B. R \cdot n}) -(R^\alpha n^\beta + R^\beta
n^\alpha)({\B. R \cdot p})] \end{eqnarray}

These are all the possible $j=1$ contributions to the symmetric, mixed
parity (inhomogeneous) structure function.

For our experimental setup II, we want to analyze the inhomogeneous
structure function in the case $\alpha = \beta = 3$, and azimuthal angle
$\phi=0$ (which corresponds to vertical separation) and we obtain the basis
tensors
\begin{eqnarray}
B_{1,1,0}^{33}(\theta) &=& \cos\theta \nonumber\\ B_{7,1,0}^{33}(\theta)
&=& 2\cos\theta \nonumber\\ B_{9,1,0}^{33}(\theta) &=& \cos^3\theta
\nonumber\\ B_{8,1,1}^{33}(\theta) &=& -2\cos\theta\sin\theta \nonumber\\
B_{1,1,-1}^{33}(\theta) &=& \sin\theta \nonumber\\ B_{9,1,-1}^{33}(\theta)
&=& \cos^2\theta\sin\theta \end{eqnarray}

Table V gives the number of free coefficients in the symmetric $j=1$ sector
in the fit to the inhomogeneous structure function for various geometrical
configurations.
\begin{table}
\begin{tabular}{|c|c|c|c|c|c|c|}
&\multicolumn{2}{c|}{$\phi= 0,\alpha=\beta=3$} & \multicolumn{2}{c|}{$\phi
= 0,\alpha=\beta=1$} & \multicolumn{2}{c|}{$\phi = 0,\alpha=3,\beta=1$}\\
\cline {2-7} $k$&$\theta \ne 0$ &$\theta = 0$ & $\theta \ne 0$ &$ \theta =
0$ & $\theta \ne 0$ & $\theta = 0$ \\ \hline
0	& 3 & 3 & 2 & 1 & 2 & 0 \\ \hline
1	& 1 & 0 & 1 & 0 & 0 & 0 \\ \hline
-1	& 2 & 0 & 3 & 0 & 2 & 1 \\ \hline \hline
Total & 6 & 3 & 6 & 1 & 4 & 1
\end{tabular}
\caption{The number of free coefficients in the symmetric $j=1$ sector for
inhomogeneous turbulence and for different geometries.} \end{table}

\end{document}